\documentstyle[prl,aps,amssymb,graphicx]{revtex}
\newcommand{\beq}{\begin{equation}}
\newcommand{\eneq}{\end{equation}}
\newcommand{\met }{\frac{1}{2}}
\newcommand{\bea }{\begin{eqnarray}}
\newcommand{\enea }{\end{eqnarray}}
\begin{document}

\tolerance 10000

%\twocolumn[\hsize\textwidth\columnwidth\hsize\csname
 %@twocolumnfalse\endcsname

\draft
\title{Hamiltonian theory of the strongly-coupled limit of the Kondo problem 
 in the overscreened case}

\author {Domenico Giuliano$~^{\dagger}$ and
         Arturo Tagliacozzo$~^{ *}$}

\address{$~^\dagger$ Dipartimento di Fisica, Universit\`a della
         Calabria, and \\
         I.N.F.N., gruppo collegato di Cosenza\\ 
         Arcavacata di Rende (CS), I-87036, Italy\\
         $~^*$ Coherentia - (INFM),
         Unit\`a di Napoli, Napoli, Italy, and\\
          Dipartimento di Scienze Fisiche Universit\`a di Napoli
         "Federico II ",\\
         Monte S.Angelo - via Cintia, I-80126 Napoli, Italy }

%\twocolumn[
\date{\today}
\maketitle
\widetext

\begin{abstract}
%\vspace*{-1.0truecm}
\begin{center}
\parbox{14cm}{By properly generalizing Nozi\`eres' Fermi liquid theory, we
construct an Hamiltonian approach to the scattering of conduction
electrons off a spin-1/2 impurity in the ovescreneed Kondo regime, as
 $T \rightarrow 0$. We derive the $S$-matrix at the interacting fixed point,
and the corresponding phase shifts, together with leading energy corrections 
to the unitary limit. We apply our results to obtain the low-temperature
dependence of the 2-channel Kondo conductance, and we relate it to 
possible transport experiments in a Quantum Dot.}

\end{center}
\end{abstract}

\pacs{
\hspace{1.9cm}
}
%]

%\narrowtext

%narrowtext

\newpage

\section{Introduction}

The Kondo Effect in metals containing magnetic impurities consists of 
an  ``anomalous'' minimum in the resistivity $\rho (T)$,
as the temperature $T $ drops  below  the ``Kondo temperature'' $T_K$. 
The minimum is due to antiferromagnetic 
scattering of conduction electrons off the localized magnetic impurities
\cite{kondo}.  Quite recently  \cite{theory,experi}, the  signature of  
Kondo interaction has  been found  in transport experiments 
across  a Quantum Dot (CD) at  Coulomb Blockade (CB). At CB, the QD is
expected  to  be  insulating, due to the  discreteness  of its levels 
\cite{experi}. Nevertheless, within a CB valley, the linear DC conductance  
may saturate at low temperatures. As $T > T_K$, the conductance 
exhibits the typical logarithmic raise \cite{silvano0}. When 
$T$  drops below  $T_K $, it saturates  to its  unitary  limit, $ G =  
2e^2 /h $. Below  $T_K$,  a perturbative   approach  in  the  coupling  is  
not feasible.

The  single  impurity  Kondo effect is  classified  according  to  the 
spin $s $  of  the impurity  and  to the  number  of  channels $\kappa $ 
of itinerant  electrons involved  in  the scattering. 
Indeed, electrons scattering off the impurity may be labelled by quantum
numbers other than the spin (for instance, angular momentum) \cite{blandin}. 
In spite of the fact that in the 
perturbative temperature region there  are  no qualitative differences 
between one-channel and many-channel effect \cite{zawa},  because 
of  the  different  nature  of  the  corresponding ground  states ($GS$), 
deep differences arise in the unitary limit, depending on $\kappa$ and $s$. 
In particular:

\begin{itemize}

\item If $\kappa = 2s $,  as $T$ is lowered  down to 0, the flow 
of the coupling strength between the impurity  magnetic  moment 
and the spin of itinerant electrons  runs all the way towards an 
infinite-coupling fixed point. At the fixed point, the impurity spin
is fully screened, and the localized magnetic moment  effectively
 disappears. The impurity rather works as a spinless scattering center
in the Fermi sea of  the itinerant electrons, and 
double occupancy at the impurity site is forbidden
(``Nozi\'eres picture'' \cite{nozieres}).
This  is  the  most  common  case: a QD with an odd number of electrons   
behaves as a spin-1/2 impurity, interacting with one channel of conduction 
electrons from the Fermi  sea  of  the  contacts.

\item If  $\kappa < 2s$, the system is ``underscreened''. A residual magnetic
moment survives on the impurity, even at $T=0$. This
moment  interacts ferromagnetically  with the spins   of the 
itinerant electrons.
$T=0$ is again an attractive infinite-coupling fixed point and, accordingly,
the system is described by Fermi Liquid (FL) theory.
An  example  of  this  case  occurs in  a two-dimensional  QD
 in  a   magnetic field at  a singlet-triplet level  crossing
\cite{eto,noi2,pustil1}.

\item If  $\kappa  > 2s$ and the  exchange  coupling  is the same for both 
channels, the impurity magnetic moment gets ``overscreened''. 
An  effective   residual magnetic moment survives on  the  impurity, 
antiferromagnetically  interacting with the spin of itinerant electrons.
As a consequence,  the infinite-coupling fixed point is now repulsive,  
as well   as  the noninteracting fixed point. The system  flows
toward an  intermediate,  finite-coupling fixed point, where
FL theory has been predicted to break down \cite{ludaff}.

The  prototype  model  for  this  case  is  the two-channel  spin 1/2 
Kondo   (2CK) model, whose low-temperature behavior will be the subject of
this paper. Our  approach   extends  Nozi\`eres' 
picture of   the  one channel Kondo  fixed  point \cite{nozieres}
to the 2CK  case.  
This  allows  us  to   highlight the Non  Fermi  Liquid (NFL) properties 
within  a  scattering  matrix   ($S-$matrix ) approach,  which  is
particularly  suitable  to studying the  conductance  across a   QD.
\end{itemize}

Overscreened Kondo effect has been hypothized to be the driving 
mechanism of low-energy physics of several systems, although, 
depending  on  the  physical  system  involved,   the  relevant 
degree of  freedom of  the  impurity may  not  be its  
spin, but  some  orbital  angular  momentum labeling 
its  energy  levels  (``orbital  Kondo'') \cite{borda}.

For instance, overscreened Kondo effect has been predicted to possibly take 
place in glassy metals \cite{zawa1}. Whether it really arises in these 
systems, or not, is still a debated question \cite{debate}.
A two channel Kondo behavior has been invoked in an experiment by Ralph and
Buhrman on clean Cu point contacts \cite{ralph}.
 A different route toward the realization of
two-channel Kondo effect in a controlled way has been recently proposed in
vertical   QD's at Coulomb blockade \cite{noi},
 or in similar mesoscopic devices\cite{gordon,glazman3}. 
According to the predictions concerning  single   impurity Kondo  models,
as $T < T_K$, the  temperature  dependent 
corrections to the conductance should   display  a crossover from  log-like 
dependence on $T$ to a dependence on $( T /T_K)^2$ for the perfectly 
screened  case, on $\sqrt{T /T_K}$ for the overscreened  case.
 In fact, the crossover to a $T^2$-dependence has been experimentally 
seen in a vertical QD with few electrons \cite{silvano0}. 
No  experimental  evidence   for  2CK in  dots has been produced  yet.

On the theoretical side, due   to  its  connections  to  
many condensed  matter problems,
the 2CK model has been studied with a number of different 
techniques \cite{chia}.  Finite-$T$ corrections have been derived by means,
 for instance,  of Bethe-ansatz like exact solutions \cite{bansatz}.
Ludwig and Affleck applied Conformal Field Theory (CFT) techniques to 
determine finite-$T$  corrections to the unitary limit, Wilson
ratios and several exact results concerning Green's functions \cite{ludaff}.
Numerical  Renormalization  Group (RG) techniques applied  to multichannel 
Kondo  have  a  long history \cite{nrg,guinea,costi1,sakai}.
Abelian bosonization  \cite{emery}  and  subsequent  refermionization
\cite{mfab} has  been  used as  well as  Majorana  fermions \cite{tsvelick}, 
 to study  the  strongly  coupled states,  by    removing    the
  unscattered  degrees  of  freedom.
 Functional  integral  methods, including   the slave  bosons technique 
\cite{woelfle} and the  Coulomb gas approach, have also been applied  to 
2CK \cite{fabr}.

A  number of approaches have been applied to describe nonequilibrium
properties such as the nonlinear conductance,
mostly  in  connection  with  the Anderson  impurity  model \cite{theory}.
These  range  from  numerical renormalization group, 
to  non-crossing approximation (NCA) \cite{meir1,kroha},
perturbative functional integral methods \cite{koenig},
perturbative renormalization group methods in real time \cite{schoeller}.

In   this  paper we study  the 2CK  model close to the  fixed  point,
by applying   bosonization  and  refermionization  of  the quantum 
particle  fields. We  consider  a   spin 1/2 impurity  at  $x=0$,  
antiferromagnetically   coupled to  a   right ($R$)  and  a left 
($L$)  one dimensional noninteracting Fermi  sea, with  an  extra  index  
$\alpha  =  1,2$,  which  labels  the   two  conduction  channels. We employ  
a scattering  approach, that is appropriate to
study  the unitary  limit  of  the conductance at  $T=0$. In particular, 
this will allow us  to calculate both fixed point conductance, and the 
leading temperature dependent corrections. By  removing  the  
degrees  of  freedom not interacting with the impurity in the unitary limit,
we  move  the  NFL  fixed  point to infinite  coupling. Accordingly, we  
apply  a perturbative strong  coupling  expansion. 
 We  first derive    the  properties  of  the  $GS$  at  the  fixed  point  
in  the  bosonic   representation  using the  lattice version of the model.
Next,  going back  to fermions  in  the  continuum  limit,  
we  obtain  a fermionic  representation   for  the 
fixed  point $S$ matrix, and get the  unitary limit
of  the  conductance,  by  using  the   Landauer  formula. 
Finally,  we derive finite $T$ corrections to the DC conductance, in 
bosonic coordinates. 
In  the  conclusions we propose a connection   between  this  
model  and  a possible  realization of the  2CK conductance  in  a QD 
\cite{noi}.

Our  representation  of the $S$ matrix  does  not  suffer  of the
 ``unitarity paradox'', since, following Ludwig and Maldacena,  we 
introduce  a  ``spin-flavor'' quantum number in  the  bosonic
  representation \cite{maldacena}. Indeed, in the unitary limit, 
the spin flavor is the only quantum number that is alleged to change 
upon scattering off the impurity. However,  some  care has  to  be  used 
when describing such a dynamics in fermionic coordinates.

The paper is organized as follows:

 In Section II we introduce the  Hamiltonian with  linearized 
bands  close  to   the  Fermi points   and the  bosonic  
representation  of  the  relevant quantum fields. 
Because  of  the  redundancy, due  to  the spin-flavor field, 
 we define two different bosonic
representations for the same fermionic field, which we refer to as $I$ and 
$II$.

 In Section III we construct the fixed  point  impurity  $GS$ by reformulating
in bosonic coordinates the  regularization
scheme proposed in Ref.\cite{tsvelick}.

 In Section IV we go  back  to  the  fermionic  representation.
Once we have identified the  physical  states, we      
implement  Nozi\`eres' scheme, by  deriving a one-body  potential  
for  the  fixed  point  fermionic  Hamiltonian. This rephrases in  
fermionic coordinates  the  spin-flavor  bosonic  field dynamics  due to
the  scattering off the  impurity. Next, we use the
Hamiltonian  we derive, to calculate the Green's functions at the fixed point. 
The one-particle Green's functions  provide us the $S$ matrix elements
in  the $I$,$II$  representation.
Our approach gives the correct result for the phase shift for
each fermion field, given by $\pm \pi / 4$. 

 In Section V and VI we  employ a Schrieffer-Wolff-like  transformation, to 
derive the  corrections to the fixed point
Hamiltonian. In particular, in Section V we show that the first correction, 
although irrelevant for what concerns the fixed point dynamics, 
selects the appropriate physical states at any point. 
 The  unitary  limit for  the  
conductance  follows  immediately, provided the degrees of freedom are
properly counted.
In Section VI  we derive  the  first irrelevant  operator, giving an  energy  
dependent correction  to the  phase shifts in  the $S-$matrix and,  
consequently,  a $T$-dependent  correction  to the  conductance. 

 In Section VII  we summarize   our conclusions  and  relate  our  results
  to the existing experimental  quest for  2CK in  QD's. 

Mathematical details of the derivation are reported in appendices
A,B,C, and D.

\section{The two-channel Kondo Hamiltonian: low-energy Fermion modes and 
bosonization} 

In this Section we introduce the model Hamiltonian for lead electrons in
 the overscreened Kondo effect. Since, in the following, we will need the 
lattice version of the Hamiltonian, we start by introducing  
the lattice version of the theory. We will propose the fermionic and the 
bosonic version of the model Hamiltonian.
 
On a system of size $L$, the lattice kinetic energy in Fourier space is 
given by:

\beq
H_T = \sum_{ k_\ell ; \sigma \alpha } [ \mu - 4 t \cos( k_\ell a ) ]
c_{\sigma \alpha}^\dagger ( k_\ell ) c_{\sigma \alpha } ( k_\ell )
\label{ola2}
\eneq
\noindent
(the $c$'s are fermionic operator in momentum space. $k_\ell = 2 \pi \ell / 
L \; ; \ell = 0 \ldots N-1 $, $\mu$ is the chemical 
potential, $a$ is the lattice step, $\alpha$ is the channel index). 

In the long wavelegth limit, expanding about the two Fermi points, one gets
the effective Hamiltonian:

\beq
H_T = - i v_f \int \; d x \sum_{\sigma \alpha} 
\biggl\{ \phi_{R ; \sigma \alpha}^\dagger ( x ) \frac{d}{ d x } 
\phi_{ R ; \sigma \alpha } ( x ) 
-   \phi_{L ; \sigma \alpha}^\dagger ( x ) \frac{d}{ d x } 
\phi_{ L ; \sigma \alpha } ( x )  \biggr\} \;\;\; , 
\label{ola5}
\eneq
\noindent
where $v_f = 4 t a \sin ( a k_F^+ )/ 2 \pi$, and:

\beq
\phi_{ L / R , \sigma \alpha } ( x ) =  
\int d p e^{ i p x } \phi_{ L/R , \sigma \alpha} ( p ) \;\;\; . 
\label{allright}
\eneq
\noindent

The isotropic lattice Kondo interaction Hamiltonian, is given by:

\beq
H_{K}^{2CK} = J {\bf S}_d \cdot [ \vec{\sigma}_1 ( 0 )
+ \vec{\sigma}_2 ( 0 ) ] \;\;\; , 
\label{eq31}
\eneq
\noindent
where $\vec{\sigma}_\alpha ( x ) = \frac{1}{2} \sum_{\sigma \sigma^{'} , 
\alpha} c_{\sigma \alpha}^\dagger (x ) \vec{ \tau }_{\sigma \sigma^{'}}
c_{ \sigma^{'} \alpha} ( x )$ and  ${\bf S}_d $ is the spin 1/2 impurity 
located  at  $x=0$.

By using  the linear combinations:

\[
\phi_{e ; \sigma \alpha} ( x ) = \frac{1}{ \sqrt{2}} [ 
\phi_{ R ; \sigma \alpha} ( x ) + \phi_{ L ; \sigma \alpha} ( - x ) ]
\]

\beq
\phi_{ o ; \sigma \alpha} ( x ) = \frac{1}{ \sqrt{2}} [
\phi_{ R ; \sigma \alpha } ( x ) - \phi_{ L ; \sigma \alpha} ( - x ) ]
\;\;\; , 
\label{ola6}
\eneq
\noindent

 $H_K^{2CK}$ takes the form:

\beq
H_K^{2CK} = J {\bf S}_d \cdot [  \vec{\sigma}_{e; 1} ( 0 )
+ \vec{\sigma}_{ e ; 2} ( 0 ) ] \;\;\; , 
\label{ola7}
\eneq
\noindent
where:

\[
\vec{\sigma}_{ e ; \alpha } ( x )  = \frac{1}{ 2 } \sum_{ \sigma \sigma^{'}} 
\phi_{ e ; \sigma  \alpha}^\dagger ( x ) \vec{\tau}_{\sigma \sigma^{'}}
\phi_{ e ; \sigma^{'} \alpha } ( x ) \;\;\; , 
\]
\noindent
that is, only the ``$e$''-fields enter the Kondo interaction Hamiltonian.
As a consequence,  we may study the Kondo dynamics by taking into account 
only the chiral fields $\phi_{ e ; \sigma \alpha}$.
From now on, we will drop the suffix $e$ from the various field operators.

In order to properly deal with the interacting fields, we bosonize 
$\phi_{ \sigma \alpha}$ \cite{ludaff}. Since we have four independent
fermionic fields, we need the same number of independent bosonic fields,
$\Psi_{ \sigma \alpha}$. Therefore, following \cite{maldacena}, we define:

\beq
\phi_{ \sigma  \alpha} ( x ) = \eta_{\sigma \alpha} : e^{- i \Psi_{\sigma 
\alpha} ( x )} : \;\;\; ,
\eneq
\noindent
where $\eta_{\sigma \alpha}$ are real Klein factors, obeying the anticommutator
algebra $\{ \eta_{\sigma \alpha} , \eta_{\tau \beta } \} = 
\delta_{\sigma \tau} \delta_{ \alpha \beta }$.

It is possible to introduce the densities of physical quantities starting
from the linear combinations \cite{maldacena,mfab}:

\beq
\Psi_{\rm ch} (x) =  \sum_{\alpha \sigma} \Psi_{\alpha \sigma} ( x )
\: \: ; \: \:
\Psi_{\rm sp} ( x ) = \sum_{\alpha \sigma} \sigma  \Psi_{\alpha \sigma} ( x )
\:\: ; \:\:
\Psi_{\rm fl} ( x ) = \sum_{\alpha \sigma} \alpha \Psi_{\alpha \sigma} ( x )
\label{eq37}
\eneq
\noindent
(In Eq.(\ref{eq37}) and in the following, whenever we use $\sigma$ and 
$\alpha$ as coefficients we mean $+ , - 1$, when $\sigma=\uparrow , 
\downarrow$, and $ + , - 1$, when $\alpha = 1 , 2$). 

The densities of charge, spin and flavor, $\rho_{\rm ch}$,   $\rho_{\rm sp}$,
 $\rho_{\rm fl}$, are given by:

\[
\rho_{\rm ch/sp/fl} (x) = \frac{1}{2 \pi} \frac{d}{ d x }
\Psi_{\rm ch/sp/fl} (x) \;\;\; . 
\]
\noindent
Out of the fields $\Psi_{\alpha \sigma}$, a fourth bosonic field,  the 
``spin-flavor'' field, independent of the first three ones, may be constructed,
given by:

\beq
\Psi_{\rm sf} ( x ) = \sum_{\alpha \sigma} \sigma \alpha \Psi_{\alpha \sigma}
( x ) \;\;\; .
\label{eq38}
\eneq
\noindent
The free dynamics of the bosonic fields $\Psi$'s is given by the bosonized
version of Eq.(\ref{ola5}), that is:

\beq
H_{\rm Bos} = \frac{v_f}{ 4 \pi} \int \; d x \sum_{X = {\rm ch,sp,fl,sf} }
\left( \frac{ d \Psi_X}{ d x } \right)^2 \;\;\; .
\label{ciccilloolupo}
\eneq
\noindent

Notice that the spin-flavor quantum number appears to be ``redundant'',
as the state of the lead electrons is fully determined by charge,
spin and flavor.
Indeed, it is possible to realize two ``inequivalent'' representations of
the fields $\phi_{ \sigma \alpha}$ in terms of the four fields
$\Psi_X$ ($X =$ch,sp,fl,sf). The former representation,  which we will refer
to as $\phi_{ \sigma \alpha}^I$, is given by:

\beq
\phi_{ \sigma \alpha}^{I} ( x ) = \eta_{\sigma \alpha} : e^{
- \frac{i}{2} [ \Psi_{\rm ch} ( x ) + \sigma \Psi_{\rm sp} ( x )
+ \alpha \Psi_{\rm fl} ( x ) + \alpha \sigma \Psi_{\rm sf} ( x ) ] } :
\;\;\; .
\label{addi1}
\eneq
\noindent

The latter representation, instead, is defined by:

\beq
\phi_{\sigma \alpha}^{II} ( x ) = \xi_{\sigma \alpha } 
: e^{ - \frac{i}{2} [ \Psi_{\rm ch} ( x ) + \sigma \Psi_{\rm sp} ( x ) +
\alpha \Psi_{\rm fl} ( x ) - \alpha \sigma \Psi_{\rm sf} ( x ) ] } :
\;\;\; .
\label{addi1bis}
\eneq
\noindent

The ``Klein-like'' factors $\xi_{\sigma \alpha}$ are 
determined by the requirement that the fields in the 
two representations anticommute with each other. Such
a requirement is achieved upon defining:

\[
\xi_{\sigma \alpha } \equiv e^{ -i \frac{ \pi }{2} 
\tilde{N}_{\sigma \alpha } }\eta_{\sigma \alpha} \;\;\; ,
\]
\noindent

where the operators $\tilde{N}_{ \sigma \alpha}$ are given by:

\[
\tilde{N}_{\sigma \alpha } =  \int \; d x \; 
[ \rho_{\rm ch} ( x ) + \sigma \rho_{\rm sp} ( x ) + \alpha \rho_{\rm fl} ( x )
- \alpha \sigma \rho_{\rm sf} ( x ) ] 
\]
\noindent
(notice the unusual definition of the $\xi$'s, which involves non real
fermionic factors. Nevertheless, both $I$ and $II$-representations
provide perfectly legitimate fermionic fields).

In Appendix A we prove that fields within the same representations obey the 
usual anticommutation relations, while fields from different 
representations anticommute with each other, that is:

\beq
\{ \phi_{  \sigma  \alpha}^{ a } ( x  ) , \phi_{  \tau  \beta }^{
b   \dagger } ( y ) \}   = \delta^{ a b  } 
\delta_{ \sigma \tau} \delta_{\alpha \beta }
\delta ( x - y ) 
\label{addi3}
\eneq
\noindent
($a , b  = I , II$).

We now start the derivation of the effective Hamiltonian in the
unitary limit.

\section{Fixed  point  impurity  Ground  State in  bosonic coordinates}

To derive the effective theory for the spin-1/2 overscreened 
Kondo system in the unitary limit, we use the regularization scheme 
introduced in Ref.\cite{tsvelick}. Such an approach allows for 
moving the intermediate coupling fixed point towards infinite coupling. 
In particular, we will reformulate the approach used in 
\cite{tsvelick} in terms of bosonic
fields, rather than in terms of Majorana fermionic fields.
 Our formalism allows for a direct derivation 
 of the subleading, finite temperature/frequency corrections
to the fixed-point dynamics. 

In bosonic coordinates, the Kondo interaction Hamiltonian is given by:

\[
H_K^{\rm 2CK} = J \biggl \{  
S_d^+ : e^{ i \Psi_{\rm sp} ( 0 ) }: : \cos ( \Psi_{\rm sf} ( 0 ) ):
+ 
S_d^- : e^{ - i \Psi_{\rm sp} ( 0 ) }:
: \cos ( \Psi_{\rm sf} ( 0 ) ):  +
S_d^z \frac{1}{2 \pi } \frac{ d \Psi_{\rm sf} ( 0 ) }{ d x } \biggr\} 
\]

\beq
\equiv   J {\bf S}_d \cdot [ \vec{\Sigma}_A ( 0 ) +
 \vec{\Sigma}_B ( 0 ) ] \;\;\; ,
\label{eq42}
\eneq
\noindent
where the spin densities $\vec{\Sigma}_{A / B} ( x )$ are given by:

\beq
\vec{\Sigma}_{A / B}^z ( x ) =
\frac{1}{4 \pi} \frac{d}{ d x } \left [ \Psi_{\rm sp}
 + / - \Psi_{\rm sf} ] (x)  \right ]
\: \: ; \: \:
\vec{\Sigma}_{A /B}^\pm ( x ) =  \frac{1}{ \sqrt{2}}
: e^{\pm i  [ \Psi_{\rm sp} + / -  \Psi_{\rm sf} ] (x)}: \;\;\; .
\label{eq43}
\eneq
\noindent

In Appendix A we prove that both $\vec{\Sigma}_A ( x )$ and $\vec{\Sigma}_B
( x )$ are $SU(2)$ spin-1/2 operators,  and show that the 
corresponding spinors at a point $x$ are realized as:

\[
| \sigma , A \rangle_x = : e^{ i \frac{\sigma}{2} [ \Psi_{\rm sp} + 
\Psi_{\rm sf} ] (x)}: | {\rm bvac } \rangle \;\;\; ,
 \]
\noindent

and:

\beq
| \sigma , B \rangle_x = : e^{ i \frac{\sigma}{2} [ \Psi_{\rm sp} - 
\Psi_{\rm sf} ] (x)}: | {\rm bvac } \rangle \;\;\; .
\label{uffa2}
\eneq
\noindent 
The doublet $ | \sigma , A / B  \rangle_x$  provides a spinor representation 
of  the $SU(2)$ group generated by

\beq
\vec{\Sigma}_{ A / B }  = \int \; d y \; 
\vec{\Sigma}_{ A / B }  ( y ) \:\:\: .
\eneq
\noindent
Also, we obtain

\beq
\vec{\Sigma}_A | \sigma , B \rangle_x = \vec{\Sigma}_B | \sigma , A \rangle_x
= 0 \;\; ,
\label{eq49}
\eneq
\noindent

 because

\[
[ \vec{\Sigma}_A , e^{ \pm \frac{i}{ 2 } [ \Psi_{\rm sp} +
\Psi_{\rm sf }  ] (x) } ] =
[ \vec{\Sigma}_B , e^{ \pm \frac{i}{ 2 } [ \Psi_{\rm sp}  -
\Psi_{\rm sf }  ] (x) } ] = 0 \:\:\: .
\]

\noindent
Eq.(\ref{eq49}) states that, if at a point $x$ the spin density associated to 
$\vec{\Sigma}_A$ is $\neq 0 $, then, at the same point, the spin density 
associated to $\vec{\Sigma}_B$ is $= 0$, and vice versa.

Such a statement is the key argument used in
Ref.\cite{tsvelick} to argue that, within such a regularization scheme, the
finite-coupling fixed point is actually moved to an infinite-coupling
point. The argument is that, since it is not possible to have at the same
point both spin densities different from 0, it is also impossible to produce
a more-than-1/2-spin composite at the origin to overscreen the impurity spin.
Therefore, the unstable overscreened fixed point disappears and
NFL-behavior is reached at an infinitely-strongly coupled
fixed point, where the impurity spin
will be fully screened in a localized spin singlet. Such a singlet must
be formed either between ${\bf S}_d$ and $\vec{\Sigma}_A ( 0 )$, or
between ${\bf S}_d$ and $\vec{\Sigma}_B ( 0 )$. Therefore, at the
fixed point  the system can lie within either one of
the two singlets $| {\rm Sin} , A , \{ \Xi \}  \rangle $, 
$ | {\rm Sin} , B , \{ \Xi \}  \rangle$, given by:

\[
| {\rm Sin}, A , \{ \Xi \} \rangle =
 \left(\frac{2 \pi \eta}{L} 
\right)^\frac{1}{4}
\frac{1}{ \sqrt{2}} \{ | \Uparrow \rangle \otimes
| \downarrow , A , \{ \Xi \}  \rangle - | \Downarrow \rangle \otimes
| \uparrow , A , \{ \Xi \} \rangle \}
\]
\noindent

\beq
| {\rm Sin} , B \{ \Xi\} \rangle = 
 \left(\frac{2 \pi \eta}{L} 
\right)^\frac{1}{4}
\frac{1}{ \sqrt{2}} \{ | \Uparrow \rangle \otimes
| \downarrow , B , \{ \Xi \} \rangle - | \Downarrow \rangle \otimes
| \uparrow , B , \{ \Xi \} \rangle \} \:\:\: ,
\label{eq410}
\eneq
\noindent
where $\eta$ is the convergence factor (see Appendix A for details)  and  
$ | \Uparrow \rangle$, $| \Downarrow \rangle$ are the two impurity states
with opposite spin polarizations. $ | \uparrow , A/B , \{ \Xi \}\rangle$,
and $ | \downarrow , A/B ,\{ \Xi \} \rangle$ are states of conduction electrons
with an $\uparrow$ or a $\downarrow$ $A/B$ particle at $x=0$,  
respectively, the state of all the other delocalized particles (globally 
denoted by $ \{ \Xi \}$) being unspecified  for   the  time being.

Let us, now, define the operator ${\cal Q}_x$: 

\beq
{\cal Q}_x = \left( \frac{ 2 \pi \eta}{L} \right )^\frac{1}{2} 
\left [ : e^{ i \Psi_{\rm sf} ( x ) } :  + : e^{ - i \Psi_{\rm sf} ( x ) } :
\right ] \;\;\; .
\label{final3}
\eneq
\noindent
It is straightforward to verify that $[ {\cal Q}_x ]^2 = 1$, $\forall x$.
Therefore, its eigenvalues are $\pm 1$. 

Since $ [ {\cal Q}_0 , H^{\rm 2CK}_K ] = 0$,where $ {\cal Q}_0 =
  {\cal Q}_{x=0}$  one may build simultaneous eigenstates of the two operators,
 given by:

\[
|{\rm Sin} , + , \{ \Xi \} \rangle =
\frac{1}{\sqrt{2}} [  | {\rm Sin} , A , \{ \Xi \} 
\rangle + i | {\rm Sin} , B , \{ \Xi \} \rangle]
\]
\noindent
\beq
| {\rm Sin} , - , \{ \Xi \} \rangle =
\frac{1}{\sqrt{2}} [  | {\rm Sin} , A , \{ \Xi \} 
\rangle -  i | {\rm Sin} , B , \{ \Xi \} \rangle] \;\;\; .
\label{anew}
\eneq
\noindent
Notice that these states are not eigenstates of the spin-flavor.

The construction   will  be  repeated  in   the  next  Section, 
 where we switch back to the fermion representation.

\section{Fixed  point  fermionic $S-$matrix }

In this Section we reformulate the results of Section III in fermionic
coordinates. We will  eventually get to the formula for 
the appropriate scattering potential in the unitary limit. Finally, we will 
take the continuum limit of our result, and derive the single-particle
$S-$-matrix for scattering off the impurity.

In bosonic coordinates, the two different representations for the
lattice fermionic fields $c_{\sigma \alpha} (x )$, which we will refer to
as $c_{\sigma \alpha}^{I / II}  ( x )$, are given by:

\beq
c_{\sigma \alpha}^{ \dagger , I / II} ( x ) = \left( \frac{ 2 \pi \eta}{L}
\right)^\frac{1}{2} : e^{ \frac{i}{2} [ \Psi_{ \rm ch} ( x ) + 
\sigma \Psi_{\rm sp} ( x ) + \alpha \Psi_{\rm sf} ( x ) + / - 
\alpha \sigma \Psi_{\rm sf} ( x ) ] } : \;\;\; .
\label{final2}
\eneq
\noindent

In terms of $c_{\sigma \alpha}^{I / II}  ( x )$, one may write down the
relevant operators acting on one-particle states: the identity operator
at point $x$

\beq
{\bf 1}_x = \sum_{\sigma \alpha} [ c_{\sigma \alpha}^{\dagger , I} 
( x ) c_{\sigma \alpha}^I ( x ) + c_{\sigma \alpha}^{\dagger , II}
( x ) c_{\sigma \alpha}^{II} ( x ) ] \;\;\; ,
\label{final7}
\eneq
\noindent

and:

\beq
{\cal Q}_x = i \sum_{\sigma \alpha} ( \sigma  \alpha )
[ c_{\sigma \alpha}^{\dagger , I } ( x ) 
c_{\sigma \alpha}^{II} ( x ) - c_{\sigma \alpha}^{\dagger , II} ( x )
 c_{\sigma \alpha}^{ I } ( x ) ] \:  .
\label{final8}
\eneq
\noindent

At a point $x$, ${\cal Q}_x$ swaps representations $I$ and $II$ with each 
other. Moreover, since in bosonic coordinates,   at  $x=0$,   
$ [ {\cal Q}_0 , H^{\rm 2CK}_K ] = 0$, we
require the same thing to hold in fermionic coordinates.

Following Nozi\`eres' approach, we construct an effective fixed point 
Hamiltonian by introducing an infinite-strength repulsive potential scattering 
at the origin, and by making it commute with ${\cal Q}_0$. 
Therefore, it is given by:

\beq
V_{\rm fp} ={\cal {P} }_0 \left \{   \lim_{\lambda \to \infty}
 [ \lambda \sum_{\sigma \alpha ; a}  c_{\sigma
 \alpha}^{\dagger  a } ( 0 ) c_{\sigma \alpha}^a ( 0 ) ]{\cal {P} }_0
\right \} \;\;\; ,
\label{final11}
\eneq
\noindent
where ${\cal P}_0 = \frac{1}{2} [  {\bf 1}_0 +  {\cal Q}_0 ]$, and 
$\lambda$ is the strength of the interaction.

We obtain the unitary Hamiltonian by adding the lattice 
kinetic energy term to $V_{\rm fp}$. By taking the continuum limit of 
the corresponding operator, one gets
the ``Nozi\`eres like'' fixed point Hamiltonian. At finite $\lambda$, this
is given by:
\[
H_{\lambda }^{2CK} \approx \int \: dx \:
\sum_{\sigma \alpha }\biggl\{ 
\left( \begin{array}{cc}
 \phi_{\sigma  \alpha }^{ \dagger,I}(x)  &  
\phi_{\sigma  \alpha }^{ \dagger,II}(x) \end{array} \right) 
\]

\beq
\biggl [ - i v_f \frac{d}{dx}{\cdot }\left( 
\begin{array}{cc}
1 & 0 \\ 
0 & 1
\end{array}
\right) + \lambda  \delta (x){\cdot }\left( 
\begin{array}{cc}
1 &  - i ( \sigma \alpha)   \\ 
 i ( \sigma \alpha )  & 1
\end{array}
\right) \biggr ]\left( 
\begin{array}{c}
\phi_{\sigma  \alpha }^{I }(x) \\ 
\phi_{\sigma  \alpha }^{II}(x)
\end{array}
\right) \biggr \} \;\;\; . 
 \label{2cham}
\eneq

\begin{figure}
\begin{center}
\includegraphics*[width=0.40\linewidth]{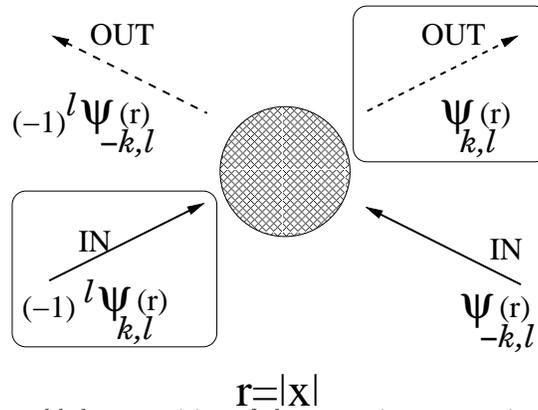}
\caption[]{Pictorial  sketch of  the even-odd 
decomposition of the scattering
process in the $(x,t)$ plane \cite{nota0}.  The  heavy dot  at the center  
is the scattering center  located  at   $x=0$.
Ingoing  and  outgoing   single  particle  wavefunctions  are shown.
One of the chiral components  in the  scattering  has  been
boxed.}
\label{figo}
\end{center}
 \end{figure}

This  Hamiltonian  envisages  a scattering process for  the    even 
 component  
of  the  field,  which  is  represented  pictorially by  the  boxed  
field  in  Fig.(\ref{figo}). 

At  $x=0 $  the projected scattering potential exchanges  the  two  
representations. Accordingly, $H_{\lambda }^{2CK}$ is  
diagonal  in  the  representation index everywhere, except at the origin.
In the following, we  use $H_{\lambda }^{2CK}$ to derive the fixed point 
one-particle Green's functions. From the Green's functions we derive the
fixed point $S$ matrix, which we will eventually use to compute the
fixed point conductance. 

Let us write down the  equations of motion for imaginary time 
one-particle Green's functions,  derived from the chiral Hamiltonian  
of  Eq.(\ref{2cham}). For simplicity, here we  just consider 
$\sigma  =\uparrow $ and  $\alpha =1 $. The Green's
functions   are defined  as $(a,b =I,II)$:

\beq
G_{ \uparrow 1  }^{a,b} ( x , \tau ; x^{'} , 
\tau^{'} ) = 
\theta ( \tau - \tau^{'} ) {\rm Tr} \biggl[ \frac{
e^{ - \beta H}}{\cal Z} \phi_{ \uparrow 1}^a ( x , \tau ) 
{\phi_{ \uparrow 1 }}^{\dagger,b} ( x^{'} , \tau^{'} ) \biggr]
- \theta (  \tau^{'} - \tau ) {\rm Tr} \biggl[ \frac{
e^{ - \beta H}}{\cal Z} {\phi_{ \uparrow 1 }}^{\dagger,b} ( x^{'} , \tau^{'} ) 
\phi_{ \uparrow 1}^a ( x , \tau ) \biggr] \;\;\; ,
\label{gree1}
\eneq
\noindent
where $\beta = 1 / k_B T$ and ${\cal Z}$ is the partition function.

The coupled 
equations of motion for $G_{\uparrow 1}^{I,I}$ and $G_{\uparrow 1}^{II , I}$
  read:
\beq
\biggl( \frac{ \partial}{ \partial \tau} + i v_f \frac{ \partial}{ 
\partial x} \biggr) G_{\uparrow 1 }^{I,I} ( x , \tau ; x^{'} , 
\tau^{'} ) = \delta ( \tau - \tau^{'} ) \delta ( x - x^{' }) 
-  \lambda  \delta ( x )   [  G_{ \uparrow 1 }^{I,I} 
( x , \tau ; x^{'} , \tau^{'} ) + i  G_{ \uparrow 1 }^{II,I} 
( x , \tau ; x^{'} , \tau^{'} )] \;\;\; ,
\label{green2}
\eneq
\noindent
and:
\beq
\biggl( \frac{ \partial}{ \partial \tau} + i v_f \frac{ \partial}{ 
\partial x} \biggr)  G_{ \uparrow 1 }^{II,I} ( x , \tau ; x^{'} , 
\tau^{'} ) =
-  \lambda  \delta ( x )  [    G_{ \uparrow 1 }^{II,I} ( x , \tau ; 
x^{'} , \tau^{'} ) - i   G_{ \uparrow 1 }^{I,I} ( x , \tau ; x^{'} , 
\tau^{'} ) ] \;\;\; .
\label{green2bis}
\eneq
\noindent

Eqs.(\ref{green2},\ref{green2bis}) have to be supplemented with the set of
equations of motion for $G_{ \uparrow 1 }^{II,II}$ and for 
$G_{ \uparrow 1}^{I,II}$, with the replacement $ i \to - i$.

The detailed solution of Eqs.(\ref{green2},\ref{green2bis}) 
is provided in Appendix B.
The $S$-matrix will come out to be diagonal with respect to
any index but the representation index,  according  to  the definition: 

\beq
G^{a,b}_{ \sigma \alpha} ( i \omega_m ;  x > 0 >  x^{'} ) = \sum_c
S^{a,c}_{\sigma \alpha} ( i \omega_m ) 
G^{c,b}_{ \sigma \alpha} ( i \omega_m ;   0 > x >  x^{'} )   \:\: ,
\label{greenkks}
\eneq
\noindent
where $\omega _{m}$ are the Matsubara frequencies. 

From the results of Appendix B, we  find:  
\[
S^{I,I}_{\uparrow 1}(i\omega _{m})=1+2\pi i\frac{
\lambda }{1+2 \lambda  {\cal F}(i\omega _{m})}
\]
\begin{equation}
S^{II,I}_{\uparrow 1}(i\omega _{m})= 2\pi \frac{\lambda 
}{1+2\lambda  {\cal F}(i\omega _{m})}  
\label{twentysix}
\end{equation}
\noindent
\begin{equation}
S^{II,II}_{\uparrow 1}(i\omega _{m})=S^{I,I}_{\uparrow 1}
(i\omega _{m})\:\:\: ,\:\:\:\:\:\:S^{I,II}_{\uparrow 1}(i\omega
_{m})=-S^{II,I}_{\uparrow 1}(i\omega _{m}) \;\;\; .
 \label{twentyeight}
\end{equation}
 Going back to real time, we get:
\[
{\cal F}(i\omega _{m}\rightarrow \omega +i\eta )=-i\pi
\biggl ( 1 - \frac{i}{\pi} \ln \biggl[\frac{
D-\omega }{D+\omega }\biggr] \biggr ) \;\;\; ,
\]
where $D$  is  a band  cutoff  energy.  

Finally :
\[
S^{I,I}_{\uparrow 1}(\omega )=S^{II,II}_{\uparrow 1}(\omega )=\frac{1-
2
\lambda \ln \biggl [\frac{D-\omega }{D+\omega }\biggr]}{1-2\pi
i \lambda  -2  \lambda  \ln \biggl[\frac{D-\omega }{D+\omega }\biggr]} 
\;\;\; ,
\]
\begin{equation}
S^{II,I}_{\uparrow 1}(\omega )=-S^{I,II}_{\uparrow 1}(\omega )=\frac{ 2\pi
\lambda }{1- 2\pi i \lambda  -
2 \lambda  \ln \biggl[\frac{D-\omega }{D+\omega }
\biggr]} \;\;\; . 
 \label{twentyseven}
\end{equation}
\noindent
This  $S-$matrix  is perfectly unitary  in   the representation  space.
the unitary limit is achieved with $\lambda  \rightarrow \infty $.
In  this  limit, the  on-shell $S$ matrix   in  the $(I,II)$ space, 
${\bf S}_{\sigma \alpha}(\omega =0)$,  is given by:
\begin{equation}
{\bf S}_{\sigma \alpha}{\bf (}\omega =0)=\left( 
\begin{array}{cc}
0 & -i (\sigma \alpha )\\ 
i ( \sigma \alpha ) & 0
\end{array}
\right)  \:\:  . 
 \label{mats}
\end{equation}

According to the  definition of phase shift $\delta$ for elastic  scattering, 
$  S  =  e^{2i\delta } $, we obtain the corresponding phase shift in
the various channels, in the unitarity limit, given by
$\delta ^{I\text{ }II}_{\uparrow 1}{\bf (}\omega =0)=-\frac{\pi }{4}$ ,
 $\delta ^{II\text{ }I}_{\uparrow 1}{\bf (}\omega =0)=\frac{\pi }{4}$. In the 
one channel spin$%
-1/2$ Kondo effect, the phase shift is $\delta_{\sigma}=\frac{\pi }{2}$.

The appearance of  matrix elements that are off-diagonal in the representation
index  avoids the unitarity paradox, at the price of alleging
scattering processes swapping the representation, corresponding to changing 
the spin-flavor quantum number by $\pm 2$. 

The derivation  above provides also the  boundary  
conditions for   the  bosonic  fields  $\Psi _X $ ($X = 
\rm{ ch, sp, fl, sf} $) at  the  origin. As $x$ crosses the location of the 
scattering center ($x=0$), none of the field changes, but $\Psi_{\rm sf}$,
according to:
\beq
\Psi_{\rm sf} ( x ) \to - \Psi_{\rm sf} ( x ) + \pi \;\;\; .
\label{green21}
\eneq
\noindent

In the following Sections, we abandon the $I,II$-representation, and 
define  the  physical  conduction states  
in fermionic coordinates. To do so, we must derive the irrelevant operators 
providing the leading corrections to the $T$=0 limit of the scattering 
dynamics.

\section{The first irrelevant correction to the fixed-point Hamiltonian,
physical   states  and  the  unitary  limit  of  the  conductance}

In  the  previous  Sections, when bosonizing the fermionic 
fields, we stressed the representation 
redundancy associated to the spin-flavor quantum number. Such a 
redundancy does not affect the physical validity of the bosonization 
procedure, as both representations possess the same observable 
quantum numbers. However, it must be taken care  of, somehow, when 
characterizing the fermionic Fock space for delocalized particles.
To  recognize  what  the  physical  fermionic  conduction  states  are,  
we  have  to  discuss the  first  irrelevant  correction  to the fixed  point 
Hamiltonian $
H_{\lambda }^{2CK} $   in  Eq.(\ref{2cham}). In order to do so, 
we will first go back to bosonic coordinates, and then derive
 the first  irrelevant  correction  to  the   bosonic   fixed  point 
Hamiltonian $H_K^{\rm 2CK}$  of  eq.(\ref{eq31}),
 by  building  the perturbation theory about
the fixed point, that is, by assuming the boundary conditions in 
Eq.(\ref{green21}), and by applying the Schrieffer-Wolff procedure to
compute the various operators. 

Irrelevant  correction
arise   when  the 
kinetic energy term is added to  $H^{\rm 2CK}_K$, and  high energy states 
with a localized triplet at the impurity 
 are alleged to take part to the scattering
process, as virtual states (each singlet state has an energy 
$ E_S = - 9 J / 4$, and  each triplet one has  energy $ E_T = 7 J / 4$,
see Appendix C for details). 

To start the derivation, let us notice that, as we show in Eq.(\ref{eq42}), 
the Kondo interaction Hamiltonian only contains spin and spin-flavor 
bosonic fields. Therefore, we may factorize out both charge and flavor fields,
and write the kinetic energy in the ``reduced'' bosonized form:
\beq
H_{\rm Red} =  \frac{ v_f}{ 4 \pi} \biggl[ \int d x \sum_{ X = {\rm sp} , 
{ \rm sf}} \biggl( \frac{ d \Psi_X }{ d x } ( x ) \biggr)^2 \biggr]
\;\;\; .
\label{additio11}
\eneq
\noindent
  This factorization   might be thought of as an artifact of the 
long-wavelength expansion, and  it may be possible that, during the 
renormalization procedure, some terms arise, coupling charge and 
flavor to the remaning degrees of freedom. However, we will assume that
such terms are irrelevant anyways, at low enough temperature.

Using the Schrieffer-Wolff procedure,
we  take as the lowest-energy  subspace the one spanned by the  singlets 
$ | {\rm Sin} , A , \{ \Xi \} \rangle$, $ | {\rm Sin} , B , \{ \Xi \} \rangle$,
and we construct an effective 
Hamiltonian as a perturbative expansion in $t^2 / J$.
 Defining  the projector onto the lowest-energy
subspace  as :
\beq
{\bf P}_0 = \sum_{u = A , B} | {\rm Sin} , u , \{ \Xi \} \rangle
\langle {\rm Sin} , u , \{ \Xi \} | \:\:\:  ,
\label{proiect1}
\eneq
\noindent 
the  effective Hamiltonian,  up  to  terms  $ {\cal{O}}( t^3 / J^2)$, is: 
\[
H_{\rm Eff} \approx {\bf P}_0  ( H_{\rm Red} + H_K ) 
 {\bf P}_0+  {\bf P}_0 \biggl\{ \frac{1}{ E_S - E_T} \biggl[ H_{\rm Red} [ {\bf 1} - {\bf P}_0 ]
H_{\rm Red} \biggr]
\]
\beq
+  \biggl( \frac{1}{ E_S - E_T} \biggr)^2
\biggl[ H_{\rm Red}  [ {\bf 1} - {\bf P}_0 ] H_{\rm Red}  [ {\bf 1} - {\bf P}_0 ] 
H_{\rm Red} \biggr] \biggr\}
{\bf P}_0 \;\;\; .
\label{additio22}
\eneq
\noindent

In  this Section  we focus  onto  the first term at the r.h.s. of 
Eq.(\ref{additio22}), ${\bf P}_0( H_{\rm Red}+H_K){\bf P}_0 $.  
In  Appendix  C, we derive the  action   of $H_{\rm Red} $   
in  the discrete  lattice model, where $H_{\rm Red}$ is substituted
by the corresponding lattice operator, $H_{T, {\rm Red}}$ (see Appendix
C for details). When computing matrix elements of $H_{T, {\rm Red}}$
between different impurity states, we 
include only the  impurity  neighboring sites (which is equivalent in
spirit to Wilson's NRG approach), and use the symbol $h_T$, when referring to
the corresponding operator. We show that, once projected on the  space  
of the singlets,  the corresponding hopping term takes the form:
 \beq
{\bf P}_0  H_{\rm Red} {\bf P}_0   \to  -\met t   {\cal Q}_0  ( {\cal Q}_{a} + 
 {\cal Q}_{-a} )    {\cal Q}_0  \;\;\; .
\label{arty}
\eneq

When written  in  terms  of fermionic  fields, the operator in Eq.(\ref{arty})
contains contributions that are off-diagonal  in  the $I,II$  representation.
However,  in the  
continuum  limit $a  \to  0$, such terms  just  add up  to  the  scattering  
potential in Eq.(\ref{2cham}),  so  that  they can accounted  for by    
substituting  $\lambda  \to  \lambda -t $. This  has  no  consequence 
on  the developements of  Section IV,  because  $\lambda \to \infty$  
at  the  end, but shows that states  $I$  and  $II$ have to be
properly mixed by hopping at any distance from the impurity. 

We recognize that the  corresponding physical  requirement is 
that $ {\cal{Q}}_x   $ commutes with the Hamiltonian not just at the origin, 
but at any point $x$. 
Hence, physical states can be constructed by using the projection operators 
given by:
\beq
{\cal {P} }_{\pm} =    \frac{1}{\sqrt{2}} \prod _x  \left ({\bf 1}_x  \pm  
{\cal Q}_x\right )
\eneq
and by requiring that pyhsical one-particle states 
are unaffected under, for instance, application of ${\cal {P} }_{+}$. 
These states are given by:
\beq
| {\rm phys} + , \sigma  \alpha \rangle_x = \frac{1}{ \sqrt{2}} 
\left [ 
c_{\sigma  \alpha }^{ \dagger, I} ( x ) - i (\sigma \alpha ) 
c_{ \sigma \alpha   }^{ \dagger,II } 
( x )  \right ] | 0 \rangle
\eneq

(The other set of one-particle physical states,  $| {\rm phys} - , \sigma  
\alpha \rangle_x $, is obtained by changing  $i \to  -i $).

Therefore, the transformation matrix ${\bf U}$, that maps 
$I,II$-representation onto 
the basis of the physical states, is:

\[
{\bf U} = \left( \begin{array}{cc} 1 / \sqrt{2}  & - i / \sqrt{2} 
\\ 1 / \sqrt{2} & i / \sqrt{2} \end{array} \right) \;\;\; .
\]
\noindent

 ${\bf U}$ transforms the ${\bf S}$ matrix  of  Eq.(\ref{mats})  as follows:

\beq
{\bf U}^\dagger \:  {\bf S}_{\sigma \alpha}  \: {\bf U} =
\left( \begin{array}{cc} -1 & 0 \\ 0 & 0 \end{array} \right)
\;\;\; .
\label{scocci}
\eneq
\noindent

As  we  have  projected  onto  just  one  of  the  two  possible sets 
of  physical  even  symmetry  states,   we find  an ${\bf S} $  matrix  for  
the  even  states which   has  just  one   non  vanishing  element (that
is -1, implying that also the even 
wavefunction has a node at the origin, as it must be
in the unitary limit). The  other possible 
 physical states decouple  from the scattering dynamics.  Had 
we chosen the other set of states, the transformed matrix would
be, instead, with only the bottom diagonal element different from 0.

At  this  point  we  have  to  recall  that this  result refers  to  the
 even-parity  wave. In fact, we get full  transmission  for   both  parities 
$l = e,o$,  because the $S$-matrix is diagonal in the basis of
physical states, and the matrix elements are given by  
${\bf S}^{(+),l}=-{\bf 1}$, for both parities $l = (e,o)$. The transmission
across the impurity is given by the  trace \cite{arturo}:

\begin{equation}
{\bf T} 
= Tr_{(+)}\left\{ \left| \frac{1}{2}\sum_{l}S^{(+),l}\right| ^{2}\right\} =
 Tr_{(+)} \{1\} \:\:\: .
\label{pseu}
\end{equation}
Here  $Tr_{(+)}$  means  tracing  over  all  physical  degrees of  freedom 
$\sigma,\alpha $.
Although the system posseses 4 degrees of freedom,  they  are  shared  by  
the two physical $\pm$-ground states.
In fact, there is  a ${\bf Z}_2 $ symmetry  breaking in  this 
system. Hence  $  Tr_{(+)} \{1\}  =  2 $. 

By applying Landauer formula to Eq.(\ref{pseu}), 
we obtain  the correct  conductivity  in  the  unitary  limit: 
\begin{equation}
\tilde{G}=\frac{e^{2}}{ h} \frac{1}{2}  
Tr_{(+)}\left\{ \left| \frac{1}{2}\sum_{l}S^{(+),l}\right| ^{2}\right\}
=\frac{2e^{2}}{h} \;\;\; .
\label{land}
\end{equation} 
This corresponds to the halving of the zero-point entropy of
the ground state, which turns to be $1/2 \ln (2)$ for each ground state.

The  fractionalization  of  the  degrees  of  freedom  can  be  understood 
also within   the   $(I,II)$ representation   for the $S-$matrix  given  
in  Eq.(\ref{mats}).  One  can  think  of  closing the  line  onto  itself 
 in  a symmetrical  ring  geometry,  with  two  equal  impurities at
opposite sites.   In  this representation, in
the  upper  branch    we  have  propagation of type-$I$ forward scattering  
into type  $II$ across  one impurity, while in the lower  branch the 
scattering is back,  $II \to I$, across the other impurity. Each process 
gives  rise  to  a phase shift  at $\omega  = 0 $, given  by  $
\delta ^e = \pm  \pi  /4 $.  The  conductance  across each  one 
impurity is $  \tilde{G}= 2 \times 2 \times \frac{e^{2}}{ h} 
  \sin ^2 \delta ^e  = 2  \frac{e^{2}}{ h}$, where 
the  first  factor  comes  from  the 
spin  degrees of  freedom, while  the  second  one comes from  the  number  
of  channels. Thus, the  unitary  limit  is again  obtained.

\section{Leading finite-$T$ corrections to  the unitary conductance}

In this Section we will explicitly write down finite-$T$ corrections to the
fixed point conductance, coming from higher-order corrections
to the fixed point Hamiltonian, (that is, from operators arising in the 
 perturbative  expansion  of  Eq.(\ref{additio22})). 
In  Appendix  C  we derive term by term  the contributions to 
Eq.(\ref{additio22}), ut ro third order.  
The  matrix  elements of  the various operators in  the 
basis of  the  singlets $ | {\rm Sin} , u , \{ \Xi \} \rangle ,\:\:
   (u =A,B )$  form  a  2$\times$2  operator matrix, acting on the two
low-energy singlets.
Eventually, we will restrict them to the subset of physical states 
defined in the previous Section.

In particular, in this Section we will derive finite-frequency 
contributions to the fermionic $S$-matrix.  These will come out to be
 $\propto \sqrt{\omega }$, which shows the NFL-nature of
the corresponding ground state \cite{ludaff}. Vertex  corrections provide
higher-order contributions, which we will not consider here.

Following the derivation of Appendix C, we see that,  to 
${\cal O} (t^2 / J)$, we obtain the  following matrix elements:

\beq
M_{AB}^{2} = M_{BA}^{2} = 0 \:\: ; \: \:
M_{AA}^{2} = M_{BB}^{2}
= \frac{2 t^2}{ E - \frac{7}{4} J } \approx - \frac{t^2}{ 2 J} \;\;\; .
\label{secondordercorrection}
\eneq
\noindent

These terms  provide just  an over-all 
trivial shift of each energy eigenvalue by a constant amount
 ${\cal{O}}(t^2)$. 

Non trivial effects, instead, arise from  the third-order corrections.

From the calculations reported in Appendix C, we see that:

\beq
M_{AA}^{3} = M_{BB}^{3} = 0 \:\: ; \:\:
 M^{3}_{AB}  = \frac{ t^3}{\pi J^2} 
\sin [ \Psi_{\rm sf} ( 0 ) ] \frac{ d \Psi_{\rm sp} ( 0 ) }{ d x }
\label{third5}
\eneq
\noindent
plus  terms ${\cal{O}}(t^2 / J )$ (and higher) that  renormalize   the  ones  
considered  before.

The third-order correction in Eq.(\ref{third5})  is not  affected 
by   the
physicality constraint  because of the operatorial relation:
$[\prod _x  {\cal Q}_x , \sin [ \Psi_{\rm sf} ( x ) ] ] = 0 $.
Since $M^{3}$ changes by $\pm 1$ the spin-flavor, the corresponding
diagonal (in $I , II$) 
contributions to the fixed-point Green's functions will be zero.
Instead, it  gives  an ${\cal{O}}(t^3/J^2) $ 
off-diagonal correction to $G_{\uparrow 1}^{I , II} $. 
In the interaction representation it reads:

\beq
 \delta G_{ \sigma \alpha , \sigma^{'}  \alpha^{'} }^{ I,II }
 ( x , \tau ; x^{'} , \tau^{'} ) 
= 
  \int_0^\beta d \tau_1 {\rm Tr} \left \{ e^{ - \beta H_0 } 
T_\tau [ \phi_{ \sigma \alpha  }^I ( x , \tau ) W_{\rm Int} ( \tau_1 ) 
\phi_{ \sigma \alpha }^{ \dagger, II  } ( x^{'} , \tau^{'} ) ]\right \}
\delta _{\sigma  \sigma^{'}  }\delta _{\alpha  \alpha^{'}  } 
\:\:\: .
\label{pertu3}
\eneq
\noindent
Computing the trace of Eq.(\ref{pertu3}) 
requires using the bosonic representation for the operators $\phi^{I,II}$,
provided in Eq.(\ref{final2}). In bosonic coordinates, $ W_{\rm Int} 
( \tau _1 )$ is  defined as:
\beq
W_{ \rm Int} ( \tau_1 ) =  \frac{t^3}{ \pi J^2} \:
  e^{ \tau_1 H_{\rm fp} } \{ : \sin [ \Psi_{ \rm sf}
( 0 ) ] : \frac{ d \Psi_{ \rm sp}}{ d x } ( 0 ) \} 
e^{ - \tau_1 H_{\rm fp} } \:\:\: . 
\label{pertu4}
\eneq
\noindent

Two independent periods appear in the Green's functions: the 
length $L$ and the inverse temperature $\beta$. A rigorous calculation
of the correlators in Eq.(\ref{pertu3}) would require the introduction
of Jacobi's elliptic $\theta$-functions. However, here we
will attempt to compute just the finite-frequency corrections, at
$T=0$. Therefore, we shall approximate Eq.(\ref{pertu3}) as:

\[
 \delta G_{ \sigma \alpha  }^{ I,II } ( x , \tau ; x^{'} , \tau^{'} ) 
= \frac{t^3}{ \pi J^2}   \: \int_0^\infty d \tau_1 
\langle : e^{ - \frac{i}{2} \Psi_{\rm ch} ( x , \tau)}: 
 : e^{  \frac{i}{2} \Psi_{\rm ch} ( x^{'}  , \tau^{'} )}: \rangle  
\langle  : e^{ - \alpha\frac{i}{2} \Psi_{\rm fl} ( x , \tau)}: 
 : e^{  \alpha 
\frac{i}{2} \Psi_{\rm fl} ( x^{'}  , \tau^{'} ) } \rangle  \times
\]

\beq
\langle  : e^{ - \sigma \frac{i}{2} \Psi_{\rm sp} ( x , \tau) } 
\frac{ \partial \Psi_{\rm sp} }{ \partial x } ( 0 , \tau_1 )  : e^{  \sigma 
\frac{i}{2} \Psi_{\rm sp} ( x^{'}  , \tau^{'} )}: \rangle 
\langle  : e^{ + \sigma \alpha \frac{i}{2} \Psi_{\rm sf} ( x , \tau)}: 
: \sin [ \Psi_{\rm sf} ( 0 , \tau _1 ) ] :   : e^{ - \sigma \alpha 
\frac{i}{2} \Psi_{\rm sf} ( x^{'}  , \tau^{'} ) } \rangle \;\;\; ,
\label{pertu5}
\eneq
\noindent
where $ \langle \ldots \rangle$ denotes ground state average.
As  shown  in    Appendix  D, for the case $\{ \sigma \alpha \} = \{  
\uparrow 1 \}$ and  in  the  limit  of  $T=0$ and large $L $, Eq.(\ref{pertu5})
provides the result:
\beq
 \delta G_{ \uparrow 1 }^{ I,II } ( i \omega _m; x , x^{'}  ) =
- \frac{ t^3}{ J^2 v_f^\frac{3}{2} } \sqrt{ \frac{ \omega}{ \pi v_f} }
e^{ i \frac{ \omega}{ v_f} ( x - x^{'} ) }
\label{pertu11}
\eneq
\noindent
with $ x-x' > 0  $. In Eq.(\ref{pertu11}), we have performed the analytic  
continuation to real frequencies for the retarded Green's function: 
$ \omega _m  \to  -i\omega  $.

Introducing  in Eq.(\ref{pertu11}) the  Kondo  temperature  as the relevant  
physical  energy  scale  according  to  the  substitution:
 $  \frac{  t^3}{ 2 \pi^{3/2} v_f J^2}  \to
\frac{1}{ \sqrt{T_K}}$, we  obtain  a leading finite-frequency correction
that goes as $\omega^\frac{1}{2}$, in agreement with the results obtained
in \cite{ludaff}. In particular, we have found a finite-$\omega$ correction
to the $S$ matrix given by:

\beq 
\delta {\bf S}_{ \sigma \alpha } ( \omega ) =  \left[ \begin{array}{cc} 0 &
i (\sigma \alpha) \sqrt{ \frac{  | \omega |}{T_K} }  \\
- i (\sigma \alpha) \sqrt{ \frac{  | \omega |}{T_K} }
& 0 \end{array} \right] \:\:\: .
\label{pertu12}
\eneq
\noindent

By projectig the result of Eq.(\ref{pertu12}) on the basis of physical
states, we get:

\beq
{\bf U}^\dagger [ {\bf S}_{\sigma \alpha} + \delta {\bf S}_{\sigma \alpha} ]
{\bf U} = \left[ \begin{array}{cc} - 1 +  \sqrt{ \frac{ | \omega | }{ T_K} } 
& 0 \\ 0 & 0 \end{array} \right] \:\:\: .
\label{emofurnimmo}
\eneq
\noindent

As  a  side  remark, let  us consider the case of  a QD  
hybridized  to  metal  contacts by   a tunneling potential $V$. If  we use 
the relations  found  within    the  Anderson  model \cite{costi},
we find the  width of  the  Kondo  resonance to be given by:
$\Gamma = \frac{4 k_B T_K}{\pi}   = 
  \pi  \nu  (0) |V |^2$,  where $V$ is the tunneling strength and $\nu ( 0 )
= 2 \pi / v_f$ is the density of states at the Fermi level.
This  implies,  in  our  case,  that 
$| V| = \frac{J^2}{t^3} \sqrt{ 2 \pi v_f^3}$.

By using the result in Eq.(\ref{emofurnimmo}), we 
compute the finite-energy transmission:

\begin{equation}
{\bf T} ( \omega , T=0 ) = 
Tr_{(+)}\left\{ \left| \frac{1}{2}\sum_{l} [ S^{(+),l} + \delta S^{(+),l} ]
\right| ^{2}\right\}
=   2  \biggl( 1  -  \frac{1}{2} \biggl| \frac{ \omega}{ 
T_K} \biggr|^\frac{1}{2}   \biggr) \;\;\; .
\label{land2}
\end{equation}

Finally, in order to obtain from Eq.(\ref{land2}) finite-temperature 
dependence of the conductance, we have to recall that, at finite temperatures,
there are two  contributions  to  the conductance: 
one   arising  from  the smearing  of  the  Fermi  surface, the  other 
from  the inelastic  processes. 
In fact, for the large $U$ Anderson model, the explicit dependence of the
transport time on temperature at the Fermi energy has to be taken into
account separately \cite{costi}. This gives:

\[
\tilde{G} ( T ) = \frac{e^2}{h}
 \int d\omega \left( - \frac{\partial f(\omega )}{\partial \omega }
\right) {\bf T} ( \omega , T )  = \frac{2 e^2}{h}
\beta
 \int_{ - \mu}^\infty \: d \omega \: \frac{ e^{ \beta \omega} }{ 
( 1 + e^{ \beta \omega} )^2 } \biggl [ 1 -  \frac{1}{2} \biggl |
\frac{ \omega}{T_K} \biggr |^\frac{1}{2} \biggr ]
+  {\bf \Delta}  (\mu ,T) 
\]

\beq
 \approx \frac{ 2 e^2}{h} \biggl[ 1 - \sqrt{ \frac{ 3 \pi }{ 8 } }
\sqrt{ \frac{ T}{T_K} } 
\biggr]
\label{condoleeza}
\eneq

where $f ( \omega )$ is Fermi distribution, and ${\bf \Delta }$ is the
total contribution coming from inelastic processes, which we neglect here,
as it is assumed to provide corrections that are higher order than
$\sqrt{T}$. 
Eq.(\ref{condoleeza}) contains the ultimate result of our derivation: the
calculation of the fixed-point contribution to the conductance, together
with the leading finite-$T$ correction, and the elucidation of the 
connection between this correction and the various scattering processes that
take place at the fixed point. As we already mentioned, our formalism 
allows also for calculating inelastic  term arising  from  
vertex  corrections, but we will not consider them here.

\section{Conclusions}

In this paper, we use  Landauer  formula to  derive the conductance at 
the   two channel  spin 1/2  overscreened Kondo fixed point, together 
with  leading finite-temperature  corrections. We perform our calculations 
within the framework of our simple Hamiltonian theory, in which  we  derive  
a suitable  fermionic representation  of  the  $S-$matrix, by  using  
bosonization  as an intermediate  step. In order for us to achieve such
 apparently simple results, we had to go through  various
mathematical approaches,  which are  used in the literature to
analyze different aspects of the overscreened Kondo problem 
\cite{ludaff,tsvelick,maldacena}.
For instance, we had to complement the regularization scheme introduced
in Ref.\cite{tsvelick} with the bosononizaton technique widely used in
\cite{ludaff}, and with the careful discussion about the role of the
spin-flavor quantum number in Ref.\cite{maldacena}.

Our research is motivated by the renewed  interest  in  the   Kondo  model,
recently arisen in connection to  conductance  experiments   across
QD's. In general, using simple models for correlated electros, like
the Anderson  model, allows  for  grasping  the
physics  involved  in tunneling  experiments across confined areas 
between two contacts, as  well as across Coulomb  blockaded  systems
(like the QD device we have in mind).
Recently, efforts  have  been  made  to  achieve an exact description  of
transport also  in  the  nonequilibrium  case, starting  from  the
integrability  of  the  two-lead  Anderson
model \cite{konik}, which  confirms previous numerical RG  results
\cite{costi1}. Despite the exactness of these results, they are unsuitable to
our case, as they  refer to  the Fermi  liquid 1CK  fixed  point.

Recently, various groups have been predicting that 2CK  could  be realized 
in  QD systems. In particular, we have proposed that  orbital 
2CK effect may arise in  a  vertical  structure with  cylindrical symmetry  
around an applied   magnetic  field \cite{noi}.

In our proposal, the dynamical degrees of freedom involved in the scattering 
across the  interacting Dot are provided by an appropriate  combination of   
the transverse angular  momentum  of lead electrons, $m$,  and of their
spin, $\sigma $. The single  particle  wavefunction of  the  delocalized
electrons  traveling  along  the $z$  direction has an orbital part factorized
in  the cross-sectional  plane. In  a linearized  band
picture and close  to  the  Fermi  energy   $\epsilon _F $,  their
wavefunction of energy  $\epsilon $  and  parity $l$  w.r.to   $z=0$  is :
\beq
\psi _{\epsilon \sim \epsilon _F ,q,m,\sigma }^l (\rho,\theta;z ) =
\left ( sign \:\:z \right )^l \: e^{i(k_F + q ) |z| }
 \varphi  _{\delta \approx 0 ,m } (\rho ) e^{im  \theta  } \:
  \chi _\sigma
\eneq
where $\delta = \epsilon -\hbar v_F |q| $, and  $\chi _\sigma $  is  the
spin wavefunction.

Another  proposal is based on the use of an  additional   lateral  QD 
to  tune  the exchange coupling to the channels to the isotropic point 
\cite{gordon}. In such a geometry, however, even  a small  anisotropy  
in  the coupling  to  the  channels  is  enough  to drive, at low enough
$T$,   the  system  from the 2CK fixed point to a 1CK  channel fixed
point,  with strong  coupling only in  the dominant channel \cite{pang}. This
 problem does  not  arise in \cite{noi}, because of  the  assumed cylindrical
symmetry.  This  symmetry    does  not  allow  for any  off-diagonal
coupling  mixing the two channels, because  it enforces  the angular  
momentum  selection  rule in  the  cross plane. Of  course,   it   is  very  
demanding  to  produce  such  a strictly cylindrical  system experimentally.

These  proposals  have  triggered a renewed  interest  in  2CK  transport.
On the theoretical side, reconsideration  of  the  matter is  relevant, 
in view of the particular kind of devices involved in the proposed experiments.
Indeed, the  scattering  approach  used in Ref.\cite{ludaff}, in conjunction
with  CFT techniques,  is better  suited for the  single
 impurity   $s-$wave  scattering in a three-dimensional  medium than  for
a  two-lead  device. In \cite{glazman3}, the authors identify the 
two-channel Kondo fixed point as a quantum critical point between two Fermi 
liquid phases, and derive its dependance  on external  parameters such  as  
temperature, magnetic  field and voltage bias accordingly.

For the purpose of understanding the physics of the transport across
mesoscopic devices, we believe our approach to be a straight connection
between the physics of the overscreened Kondo problem at the fixed point
and its physical consequences. In a clear and easy-to-follow framework
that  extends  Nozi\`eres' hypothesis,
it sheds light on the physical processes that happen at the
impurity  as $ T \to 0$, and relates them to the macroscopically detectable
non-Fermi liquid behavior in the $T$-dependence of the conductance.
The  GS at  the  NFL  fixed  point is  found  to  be  degenerate. Because
the  physical  system  only  involves  one  of  these  GS's  and  its
corresponding  excitations, the contribution
of each degree of  freedom to the conductance is  halved.
 As  discussed  in  Section IV, this  leads  to the  same  unitary  limit
  of  the  conductance  as in the one  channel  case. Finally, 
our approach might also provide an alternative route to  investigate
the  effect   of  anisotropy  on  the  conductance.

\vspace*{0.5cm}

We acknowledge fruitful discussions with M. Fabrizio, P. Wiegmann and
I. Aleiner. Work partially supported by TMR Project, contract
FMRX-CT98-0180.

\appendix

\section{Basic bosonization steps}

In this Appendix we will review some basic bosonization steps, which will 
allow us to write down fermionic fields in terms of bosonic operators.

In order to write fermionic fields in bosonic coordinates, let us
introduce a massless scalar bosonic field $\Psi (x)$, given by:

\beq
\Psi ( x ) = q + \frac{ 2 \pi}{ L } p x + i \sum_{ n \neq 0}
\frac{ \alpha_n}{n} e^{ - \frac{ 2 \pi i }{L} ( n x - i |n|
\frac{\eta}{2} ) }
\label{appe24}
\eneq
\noindent
where $\eta \equiv 0^+$ is a regularizator.

The algebra of the bosonic modes is:

\beq
[ q , p ] = i \;\;\; ; \;\;
[ \alpha_n , \alpha_m ] = n \delta_{n+m , 0} \:\:\: .
\label{appe25}
\eneq
\noindent
Therefore, we may split $\Psi (x)$ into a creation and an annihilation part,
$\Psi ( x ) = \Psi_+ ( x ) +  \Psi_- ( x )$, with:

\beq
\Psi_+ ( x ) =
q - i \sum_{n = 1}^\infty \frac{ \alpha_{-n}}{n}
e^{  \frac{ 2 \pi i n}{L} x - \frac{ \pi n }{L} \eta} 
\;\;\; ; \;\;
\Psi_- ( x ) =
\frac{  2 \pi x}{L} p + i \sum_{n = 1}^\infty \frac{ \alpha_{n}}{n}
e^{  \frac{ - 2 \pi i n}{L} x  - \frac{ \pi n }{L} \eta }
\label{appe26}
\eneq
\noindent
and:

\beq
[ \Psi_- ( y ) , \Psi_+ ( x ) ] =  - \ln (e^{  \frac{ 2 \pi i n}{L} y} - e^{  \frac{ 2 \pi i n}{L}
(x + i \eta )} ) \:\:\: .
\label{appe27}
\eneq
\noindent

The ground state of the Fock space spanned by the bosonic modes,
$ | {\rm bvac} \rangle $, is defined by:

\beq
p  | {\rm bvac}  \rangle = \alpha_n | {\rm bvac}  \rangle = 0 \;\;
(n > 0 ) \:\:\: .
\label{appe28}
\eneq
\noindent

A  fermionic field may be defined as:

\beq
c^\dagger ( x ) = : e^{i \Psi ( x ) }:
\label{appe29}
\eneq
\noindent
where the columns : : denote normal ordering with respect to
$|{\rm bvac} \rangle $.

Indeed, by using the general identity

\[
e^A e^B = e^B e^A e^{ [ A , B ] } \;\; ,
\]
\noindent
which holds if $[ A , B ]$ is a number, one gets:

\beq
 : e^{i \alpha \Psi ( x ) }: : e^{- i \beta \Psi ( y ) }: =
 e^{ i [ \alpha \Psi_+ ( x ) - \beta  \Psi_+ ( y ) ] }
 e^{ i [ \alpha \Psi_- ( x ) - \beta  \Psi_- ( y ) ] }
\left[ \frac{1}{ ( e^{ \frac{ 2 \pi i}{L} x } -
 e^{ \frac{ 2 \pi i}{L} (y + i \eta ) } )^{\alpha \beta}} \right]
\:\:\: .
\label{appe31}
\eneq
\noindent

As $\alpha , \beta = \pm 1$ and $x \neq y$, we get:

\beq
c ( x ) c ( y ) + c ( y ) c ( x ) =
c ( x ) c^\dagger ( y ) + c^\dagger ( y ) c ( x ) = 0 \:\:\: .
\label{appe32}
\eneq
\noindent

On the other hand, by properly regularizing the anticommutator we get, as
$x \to y$:

while as $x \rightarrow y$ the anticommutator between
$c ( x )$ and $ c ( y )$ remains equal to 0 (as well as the
anticommutator between $c^\dagger ( x )$ and $c^\dagger ( y )$), one
gets:

\beq
c ( x ) c^\dagger ( y ) + c^\dagger ( y )  c ( x )
 = \delta \left[ \frac{x-y}{L} 
\right] \:\:\: .
\label{appe33}
\eneq
\noindent

The density operator at a point $x$ may be expressed in bosonic coordinates
by means of the ``point-splitting regularization'' as follows:

\beq
\rho ( x ) =
 \lim_{ y \rightarrow x }
: \{ :e^{ i \Psi ( y ) } :: e^{ - i \Psi ( x ) } : \}  :
=  \frac{1}{2 \pi} \frac{ d \Psi ( x )}{ d x } \:\:\: .
\label{appe35}
\eneq
\noindent

The free kinetic Hamiltonian may also be written in a bosonized form as
follows:

\beq
H_T = - i v_f \int d x : e^{ i \Psi_+ ( x ) }  e^{ i \Psi_- ( x ) }
\frac{d}{ d x }  [  e^{ - i \Psi_+ ( x ) }  e^{ - i \Psi_- ( x ) }  ] :
= \frac{ v_f  }{4  \pi} \int d x \left[ \frac{ d \Psi ( x ) }{ d x } \right]^2
\:\:\: .
\label{appe36}
\eneq
\noindent

When  fermions carry several quantum numbers (spin, flavor),
bosonizing requires first of all introducing many bosonic fields $\Psi_X$.
Also, in order to make fermions  with different  quantum numbers anticommute,
one has to introduce a ``Klein factor'' $\eta_{\sigma \alpha}$ in front of 
each bosonized field. In general, one chooses $\eta_{\sigma \alpha}$ to be a 
real  Majorana Fermion, that is, $ (\eta_{ \sigma , \alpha} )^2 = 1$  
\cite{chia}. As a consequence, fermionic operators defined as:

\beq
c_{\sigma \alpha} ( x ) = \eta_{\sigma \alpha} : e^{ - \frac{i}{2} [ 
\Psi_{\rm ch} ( x ) + \sigma \Psi_{\rm sp} ( x ) + \alpha \Psi_{\rm fl}
( x ) + \alpha \sigma \Psi_{\rm sf} ( x ) ] } : \:\:\: ; \:\:
c^\dagger_{\sigma \alpha} ( x ) = \eta_{\sigma \alpha} : e^{ \frac{i}{2} [ 
\Psi_{\rm ch} ( x ) + \sigma \Psi_{\rm sp} ( x ) + \alpha \Psi_{\rm fl}
( x ) + \alpha \sigma \Psi_{\rm sf} ( x ) ] } : 
\label{appe38}
\eneq
\noindent
anticommute with each other for different quantum numbers, as they must
do.

The following commutation relation holds:

\beq
[ \frac{1}{ 2 \pi} \frac{ d \Psi ( x ) }{ d x } , : e^{ i \alpha \Psi ( y ) } :
] = \alpha \delta ( x - y )   : e^{ i \alpha \Psi ( y ) } : \;\;\; .
\label{pippo}
\eneq
\noindent
By using only one bosonic field $\Psi (x)$ it is possible to build
an $SU(2)$ spin-current $\vec{j} ( x )$. The components of the vector 
current density are given by:

\beq
j_\pm ( x ) = \frac{1}{\sqrt{2}} 
: e^{ \pm i \sqrt{2} \Psi (x) }:
\;\; ; \;\;
j_z ( x ) = \frac{1}{2  \pi} \frac{d \Psi}{ d x } ( x )
\:\:\: .
\label{appe39}
\eneq
\noindent

Indeed,  according to Eqs.(\ref{appe31}), in the case $\alpha = \beta = 
\sqrt{2}$, and to Eq.(\ref{pippo}),  we obtain:

\beq
[ j_+ ( x ) , j_- ( y ) ] =
 i \frac{L}{2 \pi} \delta^{'} (x-y) +  \delta (x-y)
j^z (y)
\label{appe40}
\eneq
\noindent

and

\beq
[ j^z ( x ) , j^\pm ( y ) ] =
\pm  \delta (x-y)  j^\pm ( y ) \:\:\: .
\label{appe41}
\eneq
\noindent

Eqs.(\ref{appe40},\ref{appe41}) provide us with the usual $SU(2)$ affine
algebra obeyed by the spin-density operator.

This, in particular, proves that the operators $\vec{\Sigma}_A ( x )$ and
$\vec{\Sigma}_B ( x )$ we defined in Section III are $SU(2)$ spin current
operators.

The corresponding spinors at
a point $x$ may be created by acting
$ | {\rm bvac} \rangle$ 
with  $: e^{ \pm \frac{i}{2} [ \Psi_{\rm sp} + \Psi_{\rm sf} ]
(x)}:$ and  $: e^{ \pm \frac{i}{2} [ \Psi_{\rm sp} - \Psi_{\rm sf} ]
(x)}:$, respectively. Let us define:

\beq
| \sigma ; A \rangle_x \equiv [ : e^{ \sigma \frac{i}{2} [ \Psi_{\rm sp} +
\Psi_{\rm sf} ] (x)} : ] | {\rm bvac}  \rangle
\;\; ;\;\;
| \sigma ; B \rangle_x \equiv [ : e^{ \sigma \frac{i}{2} [ \Psi_{\rm sp} -
\Psi_{\rm sf} ] (x)} :] | {\rm bvac}  \rangle \:\:\: .
\label{eq44}
\eneq
\noindent

The following commutation relations hold (either the upper or the lower 
signs hold):

\beq
[ \Sigma^\pm_A ( x ) , : e^{ \pm \frac{i}{2} [ \Psi_{\rm sp} +
 \Psi_{\rm sf}] (y )  } : ] =
[ \Sigma^\pm_B ( x ) , : e^{ \pm \frac{i}{2} [ \Psi_{\rm sp}  -
 \Psi_{\rm sf} ] (y) } : ] = 0
\label{eq45}
\eneq
\noindent

and:

\[
[ \Sigma^\pm_A ( x ) , : e^{ \mp \frac{i}{2} [ \Psi_{\rm sp}  +
 \Psi_{\rm sf}  ] (y) } : ]
 = \delta ( x - y )
: e^{ \pm \frac{i}{2} [ \Psi_{\rm sp}  +  \Psi_{\rm sf}  ] (y) } : \;\;\; ;
\]
\noindent

\beq
[ \Sigma^\pm_B ( x ) , : e^{ \mp \frac{i}{2} [ \Psi_{\rm sp}  -
 \Psi_{\rm sf}  ] (y) } : ]
 = \delta ( x - y )
: e^{ \pm \frac{i}{2} [ \Psi_{\rm sp} -  \Psi_{\rm sf}  ] (y) } :
\;\;\; .
\label{46}
\eneq
\noindent

Finally:

\[
[ \Sigma^z_A ( x ) , : e^{ \pm \frac{i}{2} [ \Psi_{\rm sp}  +
 \Psi_{\rm sf}  ] (y) } : ]
 = \pm \frac{1}{2} \delta ( x - y )
: e^{ \pm \frac{i}{2} [ \Psi_{\rm sp}  +  \Psi_{\rm sf} ] (y ) } :
\;\;\; ;
\]

\beq
[ \Sigma^z_B ( x ) , : e^{ \pm \frac{i}{2} [ \Psi_{\rm sp} -
 \Psi_{\rm sf}  ] ( y ) } : ]
 = \pm \frac{1}{2}\delta ( x - y )
: e^{ \pm \frac{i}{2} [ \Psi_{\rm sp}  -  \Psi_{\rm sf}  ] (y ) } :
\;\;\; .
\label{eq47}
\eneq
\noindent

The set of equations listed above shows that the doublet $ | \sigma , A 
\rangle$  provides a spinor representation of the $SU(2)$ group generated by
$\vec{\Sigma}_A$, and that the doublet  $ | \sigma , B 
\rangle$  provides a spinor representation of the $SU(2)$ group generated by
$\vec{\Sigma}_B$.

To conclude this Appendix, let us now
prove that fermionic fields belonging to two
different representations anticommute with each other.

Let us start from the fields in the two representations expressed in
bosonic coordinates:

\beq
\phi_{ \sigma \alpha}^I ( x ) = \eta_{ \sigma \alpha} : e^{ - \frac{i}{2}
[ \Psi_{ \rm ch } ( x ) + \sigma \Psi_{ \rm sp} ( x ) + \alpha 
\Psi_{\rm fl} ( x ) + \alpha \sigma \Psi_{\rm sf} ( x ) ] } :
\label{appe21}
\eneq
\noindent
and:

\beq
\phi_{ \sigma \alpha}^{II} ( x ) = \eta_{ \sigma \alpha} : 
e^{ - \frac{\pi i }{2}
\tilde{N}_{\sigma \alpha} }   e^{ - \frac{i}{2}
[ \Psi_{ \rm ch } ( x ) + \sigma \Psi_{ \rm sp} ( x ) + \alpha 
\Psi_{\rm fl} ( x ) - \alpha \sigma \Psi_{\rm sf} ( x ) ] } : \;\;\; ,
\label{appe22}
\eneq
\noindent
where the Klein factors $\eta_{\sigma \alpha}$ are given by real Majorana
variables. 

Clearly, fields within the same representation anticommute:

\beq
\{ \phi_{\sigma \alpha}^A ( x ) , \phi_{\sigma^{'} ,   \alpha^{'} }^{A \dagger}
( y ) \} = \delta_{\sigma \sigma^{'}} \delta_{\alpha \alpha^{'}} 
\delta ( x - y  ) \:\:\: .
\label{appe23}
\eneq
\noindent

Let us, now, consider fields from the two different representations. Let us
start with two annihilation field operators (in the case $\{ \sigma \alpha \}
\neq \{ \sigma^{'} \alpha^{'} \}$) :

\[
\{ \phi_{ \sigma \alpha}^I ( x ) , \phi_{\sigma^{'} \alpha^{'} }^{II} ( y ) \}
= 
\eta_{\sigma \alpha } \eta_{\sigma^{'} \alpha^{'} } 
\{ e^{ - \frac{ i \pi}{4} [ 1 + \sigma \sigma^{'} + \alpha \alpha^{'} 
- \sigma \sigma^{'} \alpha \alpha^{'} ]} 
 [ e^{ \frac{ 2 \pi i x }{L}} -
e^{ \frac{ 2 \pi i y}{L}} ]^{ \frac{1}{4}  [ 1 + \sigma \sigma^{'} +
\alpha \alpha^{'} - \sigma \sigma^{'} \alpha \alpha^{'} ]} ]
-
\]

\[
 [ e^{ \frac{ 2 \pi i y }{L}} -
e^{ \frac{ 2 \pi i x}{L} } ]^{ \frac{1}{4}  [ 1 + \sigma \sigma^{'} +
\alpha \alpha^{'} - \sigma \sigma^{'} \alpha \alpha^{'} ]} ]
\}
 : \exp \biggl[ - \frac{i}{2} [ \Psi_{\rm ch} ( x ) + \Psi_{\rm ch} ( y )
+ \sigma \Psi_{\rm sp} ( x ) + \sigma^{'} \Psi_{\rm sp} ( y ) + 
\]

\beq
\alpha \Psi_{\rm fl} ( x ) + \alpha^{'} \Psi_{\rm fl} ( y ) 
- \alpha \sigma \Psi_{\rm sf} ( x ) - \alpha^{'} \sigma^{'} \Psi_{\rm sf} 
( y ) ]  \biggr] : 
= 0 \:\:\: .
\label{appe24bis}
\eneq
\noindent

By following the same procedure, we obtain the result:

\beq
\{ \phi_{\sigma \alpha}^{I \dagger} ( x ) , 
\phi_{\sigma^{'} \alpha^{'} }^{II \dagger} ( y ) \} 
= 0 
\label{appe24bis1}
\eneq
\noindent
and the anticommutation relations:

\[
\{ \phi_{ \sigma \alpha}^I ( x ) , \phi_{\sigma^{'} \alpha^{'} }^{II \dagger} 
( y ) \} = 
\eta_{\sigma \alpha } \eta_{\sigma^{'} \alpha^{'} } 
\{ e^{  \frac{ i \pi}{4} [ 1 + \sigma \sigma^{'} + \alpha \alpha^{'} 
- \sigma \sigma^{'} \alpha \alpha^{'} ]}
 [ e^{ \frac{ 2 \pi i x }{L} } -
e^{ \frac{ 2 \pi i y}{L} } ]^{ - \frac{1}{4}  [ 1 + \sigma \sigma^{'} +
\alpha \alpha^{'} - \sigma \sigma^{'} \alpha \alpha^{'} ]} ]
-
\]

\[
 [ e^{ \frac{ 2 \pi i y }{L} } -
e^{ \frac{ 2 \pi i x}{L} } ]^{ - \frac{1}{4}  [ 1 + \sigma \sigma^{'} +
\alpha \alpha^{'} - \sigma \sigma^{'} \alpha \alpha^{'} ]} ]
\}
 : \exp [ - \frac{i}{2} [ \Psi_{\rm ch} ( x ) -  \Psi_{\rm ch} ( y )
+ \sigma \Psi_{\rm sp} ( x ) - \sigma^{'} \Psi_{\rm sp} ( y ) + 
\]

\beq
\alpha \Psi_{\rm fl} ( x ) -  \alpha^{'} \Psi_{\rm fl} ( y ) 
- \alpha \sigma \Psi_{\rm sf} ( x ) + \alpha^{'} \sigma^{'} \Psi_{\rm sf} 
(y ) ]  ] : 
= 0 \:\:\: ,
\label{appe24ter}
\eneq
\noindent

and:

\beq
\{ \phi_{\sigma \alpha}^{I \dagger} ( x ) , 
\phi_{\sigma^{'} \alpha^{'} }^{II} ( y ) \} = 0 \;\;\; ,
\label{appe24quattuor}
\eneq
\noindent

that complete the proof of the identities used throughout the paper.

\section{Calculation of the fixed point Green's functions}

In this Appendix we will show in detail how to calculate the fixed point
Green's functions with the method of the equations of motion. Within our
framework, this will come out to be a straightforward application of
one-dimensional scattering theory.

The equations of motion for $G_{\uparrow 1}^{I,I}$ and 
$G_{\uparrow 1}^{II , I}$ are reported in Section IV and are 
given by:

\beq
\biggl( \frac{ \partial}{ \partial \tau} - i v_f \frac{ \partial}{ 
\partial x} \biggr) G_{\uparrow 1 }^{I,I} ( x , \tau ; x^{'} , 
\tau^{'} ) = \delta ( \tau - \tau^{'} ) \delta ( x - x^{' }) 
-  \lambda \delta ( x )   [  G_{ \uparrow 1 }^{I,I} 
( x , \tau ; x^{'} , \tau^{'} ) + i  G_{ \uparrow 1 }^{II,I} 
( x , \tau ; x^{'} , \tau^{'} )]
\label{appegreen2}
\eneq
\noindent

and:

\beq
\biggl( \frac{ \partial}{ \partial \tau} - i v_f \frac{ \partial}{ 
\partial x} \biggr)  G_{ \uparrow 1 }^{II,I} ( x , \tau ; x^{'} , 
\tau^{'} ) =
-  \lambda \delta ( x )  [    G_{ \uparrow 1 }^{II,I} ( x , \tau ; x^{'} , 
\tau^{'} ) - i    G_{ \uparrow 1 }^{I,I} ( x , \tau ; x^{'} , 
\tau^{'} ) ] \:\:\: .
\label{appegreen2bis}
\eneq
\noindent

To solve the set of Eqs.(\ref{appegreen2},\ref{appegreen2bis}), 
let us introduce the Green's functions in the mixed representation:

\beq
G_{ \uparrow 1 }^{ AB} ( x , \tau ; x^{'} , \tau^{'} ) = 
\frac{1}{ \beta } \sum_{ i \omega_m } \int \: d p \: 
e^{ - i \omega_m \tau} e^{ i p x } G_{ \uparrow 1 }^{ AB} ( i \omega_m , 
p ; x^{'} ) \:\:\: .
\label{appegreen4}
\eneq
\noindent

In the mixed representation, the equations of motion become:

\beq
(-i \omega_m + v_f p )    G_{ \uparrow 1 }^{I,I} ( i \omega_m , 
p ; x^{'} ) = e^{  - i p x^{'} } 
-  \lambda \int \: d q \:   [ G_{ \uparrow 1 }^{ I,I} ( i \omega_m , 
q ; x^{'} ) + i  G_{ \uparrow 1  }^{II,I} ( i \omega_m , q ; x^{'} ) ]
\label{appegreen5}
\eneq
\noindent

and:

\beq
(-i \omega_m + v_f p )    G_{ \uparrow 1 }^{II,I} ( i \omega_m , 
p ; x^{'} ) =  
-  \lambda \int \: d q \:  [ G_{ \uparrow 1 }^{II,I } ( i \omega_m , 
q ; x^{'} ) - i  G_{ \uparrow 1 }^{I,I } ( i \omega_m , q ; x^{'} ) ]
\:\:\: .
\label{appegreen6}
\eneq
\noindent

The solution to the set of Eqs.(\ref{appegreen5} , 
\ref{appegreen6}) is given by:

\beq
G_{ \uparrow 1 }^{I,I} ( i \omega_m , p ; x^{'} ) = 
\frac{ e^{ - i p x^{'}}}{ -i \omega_m + v_f p } 
- \frac{1}{- i \omega_m + v_f p } 
\biggl[ \frac{ \lambda^{'}   }{ 
1 + 2 \lambda^{'}  {\cal F} ( i \omega_m ) } \biggr] \: v_f \int \: d q \: 
\frac{ e^{  - i q x^{'}}}{ -i \omega_m  + v_f q } \biggr) 
\label{appegreen7}
\eneq
\noindent

and by:

\beq
G_{ \uparrow 1 }^{ II,I}  ( i \omega_m , p ; x^{'} ) =  
\frac{ 1}{ - i \omega_m + v_f p } 
\biggl[ \frac{ i \lambda^{'} }{ 1 + 2 \lambda^{'}  {\cal F} ( i 
\omega_m )  } \biggr] \int \: d q \: v_f \biggl( \frac{ e^{ i q x^{'}}}{ 
- i \omega_m + v_f q } \biggr) 
\label{appegreen8}
\eneq
\noindent
where $\lambda^{'} = \lambda / v_f $ and ${\cal F} ( i \omega_m )$ is defined 
as:

\beq
{\cal F} ( i \omega_m ) = v_f \int \: d p \:  \frac{1}{- i \omega_m + v_f p } 
=   \ln \biggl[ \frac{ D + i \omega_m}{-D + i \omega_m} 
\biggr] \:\:\: .
\label{appegreen9}
\eneq
\noindent

In Eq.(\ref{appegreen9}), $D$ is a high-energy band cutoff. 

In order to derive the $S$-matrix elements, in  Section IV we use the real 
space Green's functions for $x>0$ and  $x^{'} < 0$ in the limit 
$\lambda^{'} \to \infty$. In the case $\omega_m > 0$ they are

\beq
G^{I,I}_{ \uparrow 1 } ( i \omega_m ; x , x^{'} ) = \int \: d p \: e^{ i p x }
G_{\uparrow 1}^{I,I} ( i \omega_m , p ; x^{'} ) = 
\frac{ 2 \pi i }{v_f}    \theta (  \omega_m )  
 \biggl[ 
1 -  2 \pi i   \frac{1}{ 2 {\cal F} ( i \omega_m ) }
\biggr] e^{ - \frac{ \omega_m}{ v_f } ( x - x^{' } ) }
\label{appegreen12}
\eneq
\noindent

and:

\beq
G^{II,I}_{ \uparrow 1 } ( i \omega_m ; x , x^{'} ) = \int \: d p \:  
e^{ i p x } G_{\uparrow 1}^{II,I} ( i \omega_m , p ; x^{'} ) = 
\frac{ 2 \pi i }{v_f}  \theta (  \omega_m ) 
\biggl[   2 \pi i \frac{i}{ 2 {\cal F} ( i \omega_m ) }  \biggr]
e^{ - \frac{ \omega_m}{ v_f } ( x - x^{' } ) } \:\:\: .
\label{appegreen13}
\eneq
\noindent

The ``noninteracting'' Green's functions $G_{\uparrow 1}^{(0); a,b} ( i 
\omega_m ; x , x^{'} )$ for $x - x^{'} >0$ are given by:

\beq
G_{\uparrow 1}^{(0); a,b} ( i \omega_m ; x , x^{'} ) = \int \: d p \:  
\frac{ e^{ i p ( x - x^{'} )}}{ v_f p - i \omega_m} = 
\frac{ 2 \pi i }{ v_f} \theta ( \omega_m ) e^{ - \frac{ \omega_m}{v_f} ( x - 
x^{'} ) } \delta^{a,b} \:\:\: .
\label{freegreen1}
\eneq
\noindent

From Eq.(\ref{freegreen1}), we see that  
Eqs.(\ref{appegreen12},\ref{appegreen13}) take the form:

\beq
G_{\uparrow 1}^{I,I} ( i \omega_m ; x , x^{'} ) = 
G_{\uparrow 1}^{(0); I,I} ( i \omega_m ; x , x^{'} ) 
\biggl[ 1 -  2 \pi i   \frac{\lambda^{'} }{1 +  2  \lambda^{'} {\cal F} 
( i \omega_m ) } \biggr] 
\label{freegreen2}
\eneq
\noindent

and:

\beq
G_{\uparrow 1}^{I,I} ( i \omega_m ; x , x^{'} ) = 
G_{\uparrow 1}^{(0); I,I} ( i \omega_m ; x , x^{'} ) \biggl[ 2 \pi i 
\frac{ i \lambda^{'} }{ 1 +  2  \lambda^{'} {\cal F} 
( i \omega_m ) } \biggr] \:\:\: .
\label{freegreen3}
\eneq
\noindent

From Eqs.(\ref{freegreen2},\ref{freegreen3}), 
we derive the $S$ matrix elements in Section IV.

\section{Action of ${\sl H}_T$ on the various states}

In this Appendix, we show in detail how $h_T$ acts on the states arising
from hybridization of the impurity spin with
the spin of  itinerant lead electrons. This is the mathematical support
to the derivation of Section V, where we use the results of this Appendix
to derive leading corrections to the fixed-point Hamiltonian.

 The triplet states are:

\[
| {\rm Tri} , A , 1 , \{ \Xi \} \rangle = 
| \uparrow , A , \{ \Xi \} \rangle \otimes | \Uparrow \rangle
\;\; ; \;\; 
| {\rm Tri} , B , 1 , \{ \Xi \} \rangle = 
| \uparrow , B , \{ \Xi \} \rangle \otimes | \Uparrow \rangle \;\;\; ;
\]

\[
| {\rm Tri} , A , - 1 , \{ \Xi \} \rangle = 
| \downarrow , A , \{ \Xi \} \rangle \otimes | \Downarrow \rangle
\;\; ; \;\;
| {\rm Tri} , B , -  1 , \{ \Xi \} \rangle = 
| \downarrow , B , \{ \Xi \} \rangle \otimes | \Downarrow \rangle
\;\;\; ;
\]

\[
| {\rm Tri} , A , 0 , \{ \Xi \} \rangle = 
\frac{1}{ \sqrt{2}} [ | \uparrow , A , \{ \Xi \} \rangle \otimes | 
\Downarrow \rangle + | \downarrow , A , \{ \Xi \} \rangle \otimes | 
\Uparrow \rangle ] \;\;\; ;
\]

\beq
| {\rm Tri} , B , 0 , \{ \Xi \} \rangle = 
\frac{1}{ \sqrt{2}} [ | \uparrow , B , \{ \Xi \} \rangle \otimes | 
\Downarrow \rangle + | \downarrow , B , \{ \Xi \} \rangle \otimes | 
\Uparrow \rangle ] \;\;\; .
\label{trilist}
\eneq
\noindent

Let us introduce the lattice operators $b_X$ given by:

\beq
b_X ( x ) \equiv \sqrt{ \frac{ 2 \pi \eta}{ L}}  : 
e^{ - i \phi_X ( n a ) }:
\;\; ; \;
X = {\rm sp, sf} \;\;\; .
\label{additio15}
\eneq
\noindent

 To start the derivation, let us
rewrite $H_{\rm Red}$ in a lattice form, by using the operators $b_{\rm sp}$,
$b_{\rm fl}$. We get
the reduced lattice kinetic energy operator $H_{T , {\rm Red}}$, given by:

\beq
H_{T , {\rm Red}} = \sum_{x, X = {\rm sp,sf}} \{ b_X^\dagger ( x ) 
b_X ( x ) -  t [ b_X^\dagger ( x ) ( b_X ( x + a ) + b_X ( x - a ) )
+ ( b_X^\dagger ( x + a ) + b_X^\dagger ( x - a ) ) b_X ( x ) ] \}
\:\:\: .
\label{okkey1}
\eneq
\noindent

From the operator in Eq.(\ref{okkey1}), we may single out the term that
acts on the Fock spaces of the states formed by hybridization between the
spin of the localized impurity at the origin, and the spin of conduction
electrons at $x = \pm a$. Such a term is given by

\[
h_T = - t \sum_{ X = {\rm sp, sf}} [ b_X^\dagger ( 0 ) 
( b_X ( a ) + b_X ( - a )) + ( b_X^\dagger ( a ) + b_X^\dagger ( - a ) )
b_X ( 0 ) ] \:\:\: .
\]
\noindent

The action of $h_T$ on the singlets is given by:

\[
h_T |{\rm Sin} , A \rangle = 
\frac{t}{\sqrt{2}} \{ [ b_{\rm sp}( a ) + b_{\rm sp} ( - a ) ] 
| {\rm Tri} , B , 1 \rangle - [ b_{\rm sp}^\dagger ( a ) + 
b_{\rm sp}^\dagger ( - a ) ] | {\rm Tri} , B , - 1 \rangle \}
\]

\beq
- \frac{t}{2} [ b_{\rm sf} ( a ) + b_{\rm sf} ( - a ) + b_{\rm sf}^\dagger 
( a  ) + b_{\rm sf}^\dagger ( - a ) ] | {\rm Sin} , B \rangle 
+ \frac{t}{2} [ b_{\rm sf} ( a ) + b_{\rm sf} ( - a ) - b_{\rm sf}^\dagger
( a ) - b_{\rm sf}^\dagger ( - a ) ] | {\rm Tri} , B , 0 \rangle
\label{unoplus}
\eneq
\noindent

and by:

\[
h_T | {\rm Sin} , B \rangle = 
\frac{t}{\sqrt{2}} \{ [ b_{\rm sp}( a ) + b_{\rm sp} ( - a ) ] 
| {\rm Tri} , A , 1 \rangle - [ b_{\rm sp}^\dagger ( a ) + 
b_{\rm sp}^\dagger ( - a ) ] | {\rm Tri} , A , - 1 \rangle \}
\]

\beq
- \frac{t}{2} [ b_{\rm sf} ( a ) + b_{\rm sf} ( - a ) + b_{\rm sf}^\dagger 
( a  ) + b_{\rm sf}^\dagger ( - a ) ] | {\rm Sin} , A \rangle 
- \frac{t}{2} [ b_{\rm sf} ( a ) + b_{\rm sf} ( - a ) - b_{\rm sf}^\dagger
( a ) - b_{\rm sf}^\dagger ( - a ) ] | {\rm Tri} , A , 0 \rangle
\:\:\: .
\label{dueplus}
\eneq
\noindent

When acting on the triplet states, instead, $h_T$ provides the following 
results:

\beq
h_T | {\rm Tri} , A , 1 \rangle = - t [ b_{\rm sf}^\dagger ( a ) + 
b_{\rm sf}^\dagger ( - a ) ] | {\rm Tri} , B , 1 \rangle  
+  \frac{t}{\sqrt{2}} [ b_{\rm sp}^\dagger ( a ) + b_{\rm sp}^\dagger ( - a ) ]
| {\rm Sin} , B \rangle 
- \frac{t}{\sqrt{2}} [ 
 b_{\rm sp}^\dagger ( a ) + b_{\rm sp}^\dagger ( - a ) ] |{\rm Tri} , B , 0 
\rangle
\label{treplus}
\eneq
\noindent

\beq
h_T | {\rm Tri} , B , 1 \rangle = - t [b_{\rm sf} ( a ) + b_{\rm sf} ( - a ) ]
| {\rm Tri} , A , 1 \rangle
+  \frac{t}{\sqrt{2}} [ b_{\rm sp}^\dagger ( a ) + b_{\rm sp}^\dagger ( - a ) ]
| {\rm Sin} , A \rangle 
- \frac{t}{\sqrt{2}} [ 
 b_{\rm sp}^\dagger ( a ) + b_{\rm sp}^\dagger ( - a ) ] |{\rm Tri} , A , 0 
\rangle
\label{quattroplus}
\eneq
\noindent

\beq
h_T | {\rm Tri} , A ,  - 1 \rangle = - t [ b_{\rm sf} ( a ) + b_{\rm sf} 
( - a ) ] | {\rm Tri} , B , - 1 \rangle 
- \frac{t}{\sqrt{2}} [ b_{\rm sp} ( a ) + b_{\rm sp} ( - a ) ] | {\rm Sin} , 
B \rangle
 - \frac{t}{\sqrt{2}}  [ b_{\rm sp} ( a ) + b_{\rm sp} ( - a ) ]
| {\rm Tri} , B , 0 \rangle
\label{cinqueplus}
\eneq
\noindent

\beq
h_T | {\rm Tri} , B ,  - 1 \rangle = - t [ b_{\rm sf}^\dagger ( a ) + 
b_{\rm sf}^\dagger ( - a ) ] | {\rm Tri} , A , - 1 \rangle 
- \frac{t}{\sqrt{2}} [ b_{\rm sp} ( a ) + b_{\rm sp} ( - a ) ] | {\rm Sin} , 
A \rangle
 - \frac{t}{\sqrt{2}}  [ b_{\rm sp} ( a ) + b_{\rm sp} ( - a ) ]
| {\rm Tri} , A , 0 \rangle
\label{seiplus}
\eneq
\noindent

\[
h_T |{\rm Tri} , A , 0 \rangle = 
- \frac{t}{\sqrt{2}}
\{ [ b_{\rm sp} ( a ) + b_{\rm sp} ( - a ) ] |{\rm Tri} , B , 1 \rangle 
+ [ b_{\rm sp}^\dagger ( a ) + b_{\rm sp}^\dagger ( - a ) ] 
| {\rm Tri} , B , - 1 \rangle \} 
\]

\beq
- \frac{t}{2} \{ [ b_{\rm sf} ( a ) + b_{\rm sf} ( -a ) + b_{\rm sf}^\dagger
( a ) + b_{\rm sf}^\dagger ( - a ) ] \} | {\rm Tri} , B , 0 \rangle
- [ b_{\rm sf} ( a ) + b_{\rm sf} ( - a ) - b_{\rm sf}^\dagger ( a ) 
- b_{\rm sf}^\dagger ( - a ) ] |{\rm Sin} , B \rangle \}
\label{setteplus}
\eneq
\noindent

\[
h_T |{\rm Tri} , B , 0 \rangle = 
- \frac{t}{\sqrt{2}}
\{ [ b_{\rm sp} ( a ) + b_{\rm sp} ( - a ) ] |{\rm Tri} , A , 1 \rangle 
+ [ b_{\rm sp}^\dagger ( a ) + b_{\rm sp}^\dagger ( - a ) ] 
| {\rm Tri} , A , - 1 \rangle \}
\]

\beq
 - \frac{t}{2} \{ [ b_{\rm sf} ( a ) + b_{\rm sf} ( -a ) + b_{\rm sf}^\dagger
( a ) + b_{\rm sf}^\dagger ( - a ) ] \} | {\rm Tri} , A , 0 \rangle
+ [ b_{\rm sf} ( a ) + b_{\rm sf} ( - a ) - b_{\rm sf}^\dagger ( a ) 
- b_{\rm sf}^\dagger ( - a ) ] |{\rm Sin} , A \rangle \} \:\:\: .
\label{ottoplus}
\eneq
\noindent

The equations listed above are all we need, in order to work out
subleading corrections to the fixed point Hamiltonian.

We apply
the Schrieffer-Wolff transformation introduced in Eq.(\ref{additio22}), 
by including just the hopping term between sites nearest neighbors of the
impurity. The effective perturbative perturbative potential is:
\[
V_{\rm Eff} \approx {\bf P}_0 \biggl\{ ( h_{T }  + H_K ) 
+  \frac{1}{ E_S - E_T} \biggl[ h_T [ {\bf 1} - {\bf P}_0 ]
h_T \biggr]
\]
\beq
+  \biggl( \frac{1}{ E_S - E_T} \biggr)^2
\biggl[ h_T  [ {\bf 1} - {\bf P}_0 ] h_T  [ {\bf 1} - {\bf P}_0 ] 
h_T \biggr] \biggr\}
{\bf P}_0 \;\;\; .
\label{additio222}
\eneq
\noindent

The first term at the r.h.s. of Eq.(\ref{additio222}), can be expressed
in terms of the operators ${\cal Q}_0$, and ${\cal Q}_{\pm a}$, leading to
the Eq.(\ref{arty}).

The ${\cal O} (t^2 / J)$ of Eq.(\ref{additio222})
provides the following matrix elements:

\beq
M_{AB}^2 = M_{BA}^2 = 0 \;\; ; \;\;
M_{AA}^2 = M_{BB}^2 
= \frac{2 t^2}{ E - \frac{7}{4} J } \approx - \frac{t^2}{ 2 J}
\:\:\: . 
\label{secondordercorrection2}
\eneq
\noindent

Therefore, on $ | {\rm Sin} , \pm , \{ \Xi \}  \rangle$,
 second-order (in $t$) dynamics just yields an over-all 
trivial shift of each energy eigenvalue by a constant amount. 

Non trivial, effects, instead, arise from  third-order corrections. Indeed,
while we have once more:

\beq
M_{AA}^3 = M_{BB}^3 = 0 \;\; \;\; ,
\label{thirdo1}
\eneq
\noindent

on the other hand, we obtain:

\[
 ( 4 J)^2 M^3_{ AB} = 
\sum_{X X^{'}} \langle {\rm Sin} , A , \{ \Xi \}
  | h_T | E_X \rangle \langle E_X | h_T
| E_{X^{'}} \rangle \langle  E_{X^{'}} | h_T |{\rm Sin} , B \rangle = 
\]

\[
- \frac{3 t^3 }{4} ( b_{\rm sf} ( a ) + b_{\rm sf} ( - a ) + b_{\rm sf}^\dagger
( a ) + b_{\rm sf}^\dagger ( - a ) )
\]

\beq
- \frac{t^3}{4} [ ( b_{\rm sp} ( a ) + b_{\rm sp} ( - a ) ) , 
( b_{\rm sp}^\dagger ( a ) + b_{\rm sp}^\dagger ( - a ) ) ] 
( b_{\rm sf} ( a ) + b_{\rm sf} ( - a ) - b_{\rm sf}^\dagger ( a ) 
- b_{\rm sf}^\dagger ( - a )  )  \:\:\: ,
\label{quattroplusbis}
\eneq
\noindent
where $ | E_X \rangle$ is a generic high-energy triplet.

From Eq.(\ref{quattroplusbis}), we see that $M_{AB}^3$ contains a contribution
proportional to the term we found to ${\cal O} ( t )$. Therefore, it
may be accounted for by means of a slight renormalization of
$\lambda^{'}$, that is ${\cal O} ( t^3 / J^2)$ and is absolutely irrelevant. 
Since this effect is trivial, we  will not consider it here. 
The nontrivial part of the
correction operator  can be derived by recalling the
basic bosonization rules listed in Appendix A. In particular, we obtain the
following expression for the commutator in Eq.(\ref{quattroplusbis}):

\beq
[ ( b_{\rm sp}^\dagger  ( a ) + b_{\rm sp}^\dagger ( - a )) , 
( b_{\rm sp} ( a ) + b_{\rm sp} ( - a ) ) ] = 
 \frac{1}{\pi} \frac{ d \Psi_{\rm sp} ( 0 ) }{ d x } \:\:\: .
\label{cinqueplusbis}
\eneq
\noindent
 
Therefore, taking the limit $a \to 0$ in the regular part of the result in 
Eq.(\ref{quattroplusbis}), we get the result:

\beq
 M^3_{AB}  = M_3 = \frac{ t^3}{\pi J^2} 
\sin [ \Psi_{\rm sf} ( 0 ) ] \frac{ d \Psi_{\rm sp} ( 0 ) }{ d x }
\label{third55}
\eneq
\noindent
plus the irrelevant terms discussed before.

\section{Calculation of leading finite-frequency corrections to the
$S$-matrix}

In this Appendix, we will show in detail how to calculate leading 
finite-frequency corrections to the $S$-matrix. To do so, 
first of all, let us recall what are the relevant correlators we need. We
may calculate them according to the basic bosonization rules of 
Appendix A (and to the fact that all the involved fields are chiral). 
We obtain:

\beq
\langle {\bf {\cal T}}_\tau [ : e^{ - \frac{i}{2} \Psi ( x , \tau ) } : 
 : e^{  \frac{i}{2} \Psi ( x^{'} , \tau^{'} ) } : ] \rangle 
= \frac{1}{ \biggl[ e^{ \frac{2 \pi i }{L} ( x + i v_f \tau ) } - 
e^{ \frac{ 2 \pi i }{L} ( x^{'} + i v_f \tau^{'} ) } \biggr]^\frac{1}{4} }
\label{ciccicicci}
\eneq
\noindent
(this will provide us with the charge-charge and  the flavor-flavor part
of the relevant correlator).

Moreover, we get:

\beq
\langle {\bf {\cal T}}_\tau  [ : e^{ - \frac{i}{2} \Psi_{\rm sp} ( x ,\tau ) }:
\frac{ \partial \Psi_{ \rm sp} ( 0 , \tau_1 ) }{ \partial x} 
: e^{ \frac{i}{2} \Psi_{\rm sp} ( x^{'} , \tau^{'} ) } : ] \rangle =
- \frac{\pi}{L} \frac{ e^{ \frac{ 2 \pi v_f}{ L } \tau_1} [
e^{ \frac{ 2 \pi i }{L} ( x + i v_f \tau ) } - e^{ \frac{ 2 \pi i }{L} 
( x^{'} + i v_f \tau^{'} ) } ]^\frac{3}{4} }{ [ 
e^{ \frac{ - 2 \pi v_f}{ L } \tau_1} - e^{ \frac{ 2 \pi i }{L} 
( x + i v_f \tau ) } ] 
 [ e^{ \frac{ - 2 \pi v_f}{ L } \tau_1} - e^{ \frac{ 2 \pi i }{L} 
( x^{'} + i v_f \tau^{'}  ) } ] } \:\:\: .
\label{nota18}
\eneq
\noindent

Finally, we obtain:

\[
\langle {\bf {\cal T}}_\tau [ :e^{ - \frac{i}{2} \Psi_{\rm sf} ( x , \tau ) }:
: e^{ i \Psi_{\rm sf} ( 0 , \tau_1 ) } :
: e^{ - \frac{i}{2} \Psi_{\rm sf} ( x^{'} , \tau^{'} ) } : ] \rangle =
\langle {\bf {\cal T}}_\tau [ :e^{  \frac{i}{2} \Psi_{\rm sf} ( x , \tau ) }:
: e^{ - i \Psi_{\rm sf} ( 0 , \tau_1 ) } :
: e^{  \frac{i}{2} \Psi_{\rm sf} ( x^{'} , \tau^{'} ) } : ] \rangle =
\]

\beq
\frac{ [ e^{ \frac{ 2 \pi i}{L} ( x + i v_f \tau ) } - 
e^{ \frac{ 2 \pi i}{L} ( x^{'} + i v_f \tau^{'} ) } ]^\frac{1}{4} }{
[ 
e^{ \frac{ - 2 \pi v_f}{L} \tau_1 } - e^{ \frac{ 2 \pi i}{L} 
( x + i v_f \tau )} ]^\frac{1}{2} [ 
e^{ - \frac{ 2 \pi v_f}{L} \tau_1 } - e^{ \frac{ 2 \pi i}{L} ( x^{'} + i 
v_f \tau^{'} )} ]^\frac{1}{2} } \:\:\: .
\label{nota19}
\eneq
\noindent

By putting together all the correlators in Eqs.(\ref{ciccicicci},\ref{nota18},
\ref{nota19}), we obtain the integral in Eq.(\ref{pertu5}). To calculate 
the integral, we use as an auxiliary variable $\xi = \exp [ - \frac{ 2 \pi}{L} 
v_f \tau ]$. Therefore, after all the substitutions and the variable 
replacements have been made, the integral reads:

\[
\frac{ i t^3}{  2 \pi v_f J^2} \: \left( \frac{2 \pi}{L} \right)^\frac{3}{2} 
\int_0^1 \: d \xi \: \frac{ 
[ e^{ \frac{ 2 \pi i}{L} ( x + i v_f \tau ) } - 
 e^{ \frac{ 2 \pi i}{L} ( x^{'}  + i v_f \tau^{'} ) } ]^\frac{1}{2} }{
[ \xi - e^{ \frac{ 2 \pi i}{L} ( x + i v_f \tau ) } ]^\frac{3}{2} 
[ \xi - e^{ \frac{ 2 \pi i}{L} ( x^{'} + i v_f \tau^{'} ) } ]^\frac{3}{2} }
\]

\beq
= \frac{ i t^3}{ 2 \pi v_f J^2} \frac{ ( 2 \pi / L )^\frac{3}{2} }{ 
[ e^{ \frac{ 2 \pi i}{L} ( x + i v_f \tau ) } - 
 e^{ \frac{ 2 \pi i}{L} ( x^{'}  + i v_f \tau^{'} ) } ]^\frac{3}{2} }
\biggl\{ \frac{ [ e^{ \frac{ 2 \pi i}{L} ( x + i v_f \tau ) } + 
 e^{ \frac{ 2 \pi i}{L} ( x^{'}  + i v_f \tau^{'} ) } ]}{ \sqrt{ 
 e^{ \frac{ 2 \pi i}{L} ( x + i v_f \tau ) }
e^{ \frac{ 2 \pi i}{L} ( x^{'}  + i v_f \tau^{'} ) } } }
- \frac{ [ e^{ \frac{ 2 \pi i}{L} ( x + i v_f \tau ) } + 
 e^{ \frac{ 2 \pi i}{L} ( x^{'}  + i v_f \tau^{'} ) } ] - 2 }{ \sqrt{ 
( e^{ \frac{ 2 \pi i}{L} ( x + i v_f \tau ) } - 1 )(
e^{ \frac{ 2 \pi i}{L} ( x^{'}  + i v_f \tau^{'} ) } - 1 ) }  } \biggr\}
\label{nota19bis}
\eneq
\noindent
(notice the extra prefactor of $ (2 \pi / L)^\frac{3}{2}$, which we have
introduced in order to make the bosonization rules of Appendix A consistent
with the normalization for the Green's functions introduced in Appendix B).

In Section VI we point out that the integral in Eq.(\ref{nota19bis}) is
computed in the limit of very large $L , \beta $. Such a limit yields 
the ultimate approximate formula:

\beq
\frac{ i t^3}{  2 \pi v_f J^2} \biggl( \frac{1}{  i } \biggr)^\frac{3}{2} 
\frac{ 1 }{ [ x - x^{'} + i v_f \tau]^\frac{3}{2} }
\label{nota20}
\eneq
\noindent

When going back to (Matsubara) frequency space, we have to calculate:

\beq
\frac{ i t^3}{ 2 \pi v_f J^2} \biggl( \frac{1}{  i } \biggr)^\frac{3}{2} 
\int_{ - \infty}^\infty \: d \tau \: 
\frac{ e^{ i \omega_m \tau} }{ [ x - x^{'} + i v_f \tau]^\frac{3}{2} }
\:\:\: .
\label{nota20bis}
\eneq
\noindent

The integral in Eq.(\ref{nota20bis}) is calculated as follows:

\[
\frac{ i  t^3}{ 2 \pi v_f J^2} \biggl( \frac{1}{ i } \biggr)^\frac{3}{2} 
\int_{ - \infty}^\infty \: d \tau \: 
\frac{ e^{ i \omega_m \tau} }{ [ x - x^{'} + i v_f \tau]^\frac{3}{2} } 
= \frac{ -i t^3}{ \pi v_f J^2}  \biggl( \frac{1}{  i } 
\biggr)^\frac{3}{2} 
\frac{ \partial}{\partial x} \int_{ - \infty}^\infty \frac{ d s}{ \sqrt{\pi}}
e^{ - s^2 ( x - x^{'} )} 2 \pi \int_{ - \infty}^\infty \: d \tau \: e^{ i \tau
( \omega_m - s^2 v_f )} 
\]

\beq
=  - \frac{ i t^3}{J^2} \sqrt{ \frac{ \omega_m}{ \pi v_f } }
( i v_f )^{ - \frac{3}{2}}
\theta ( \omega_m ) e^{ - \frac{ \omega_m}{ v_f } ( x - x^{'} ) } 
\label{hopefinal1}
\eneq
\noindent

with $ x - x^{'} > 0$. 

When rotating back to real times/frequencies, the result in 
Eq.(\ref{hopefinal1}) reads:

\beq
- \frac{ t^3}{ J^2 v_f^\frac{3}{2} } \sqrt{ \frac{ | \omega | }{ \pi v_f} }
e^{ i \frac{ \omega}{ v_f} ( x - x^{'} ) } \:\:\: .
\label{whataball}
\eneq
\noindent

By  going  back    to real  frequency,   this provides the finite-frequency
 correction to the
${\bf S }$-matrix that we have used in Section VI to derive  the 
conductance.


\begin{thebibliography}{99}
\bibitem{nota0} For a one-dimensional  elastic channel with momentum $k$, the 
scattering  amplitudes  $f $ are defined by 
\beq
\Psi _k ( x )  = \biggr \{
\begin{array}{c}  
 e^{ikx} + f_> e^{ikx}\hspace*{0.5cm} x>>0 \\
 e^{ikx} + f_< e^{-ikx} \hspace*{0.5cm} x<<0
 \end{array} 
\nonumber
\eneq
The even and odd parity  components ($l=e$ and $l=o$), together with the 
corresponding amplitudes, are defined by:
\[
\Psi _k (x) = [ \cos (kr) + f_e e^{ikr} ] + sgn(x)[i \sin(kr) + f_o e^{ikr} ]
\equiv \psi_{k , e} (r)+  sgn(x) \psi_{k , o} (r) 
%\label{eq2}
\]
where $r \equiv |x| $ and
$f_e =\met ( f_<+f_>) $ and $f_o =\met ( f_>-f_<) $.
\end{thebibliography}
\end{document}